\documentclass[a4paper,11pt]{article}
\pdfoutput=1 

\usepackage{jheppub} 

\usepackage[T1]{fontenc} 

\usepackage{amssymb,amsmath}
\usepackage{mathtools}
\usepackage{graphicx}
\usepackage{epsfig}
\usepackage{epstopdf}
\usepackage{setspace}
\usepackage{dsfont}
\usepackage{latexsym}
\usepackage{tikz}
\usepackage{psfrag}
\usepackage{booktabs}
\usepackage{bbm}
\usepackage[toc]{appendix}
\usepackage{color}
\usepackage{datetime}
\usepackage{tensor}
\usepackage{lscape}
\usepackage{lipsum}

\usepackage{tikz}
\usetikzlibrary{arrows,shapes}
\usetikzlibrary{trees}
\usetikzlibrary{patterns}
\usetikzlibrary{matrix,arrows} 				
\usetikzlibrary{positioning}				
\usetikzlibrary{calc,through}				
\usetikzlibrary{decorations.pathreplacing}  
\usepackage{pgffor}							

\usetikzlibrary{decorations.pathmorphing}	
\usetikzlibrary{decorations.markings}
\tikzset{
    vector/.style={decorate, decoration={snake}, draw},
	provector/.style={decorate, decoration={snake,amplitude=2.5pt}, draw},
	antivector/.style={decorate, decoration={snake,amplitude=-2.5pt}, draw},
        smallvector/.style={decorate, decoration={snake,amplitude=1.5pt,post length=0.5mm}, draw},
    fermion/.style={draw=black, postaction={decorate},
        decoration={markings,mark=at position .55 with {\arrow[draw=black]{>}}}},
    fermionbar/.style={draw=black, postaction={decorate},
        decoration={markings,mark=at position .55 with {\arrow[draw=black]{<}}}},
    fermionnoarrow/.style={draw=black},
    gluon/.style={decorate, draw=black,
        decoration={coil,amplitude=4pt, segment length=5pt}},
    scalar/.style={dashed,draw=black, postaction={decorate},
        decoration={markings,mark=at position .55 with {\arrow[draw=black]{>}}}},
    scalarbar/.style={dashed,draw=black, postaction={decorate},
        decoration={markings,mark=at position .55 with {\arrow[draw=black]{<}}}},
    scalarnoarrow/.style={dashed,draw=black},
    electron/.style={draw=black, postaction={decorate},
        decoration={markings,mark=at position .55 with {\arrow[draw=black]{>}}}},
    bigvector/.style={decorate, decoration={snake,amplitude=4pt}, draw},
    arrow/.style={draw=black, postaction={decorate},
        decoration={markings,mark=at position 1 with {\arrow[draw=black]{>}}}},
}

\tikzstyle{block} = [draw, rectangle, 
    minimum height=3em, minimum width=6earticlem]

\definecolor{darkblue}{rgb}{0.2, 0, 0.8}

\numberwithin{equation}{section}

\def\be{\begin{equation}}
\def\ee{\end{equation}}
\def\bea{\begin{eqnarray}}
\def\eea{\end{eqnarray}}
\def\ba{\begin{array}}
\def\ea{\end{array}}
\def\bd{\begin{displaymath}}
\def\ed{\end{displaymath}}

\def\>{\rangle} 
\def\<{\langle} 
\def\Dsl{D \hskip-.6em \raise1pt\hbox{$ / $ } }
\def\to{\rightarrow}

\title{Constraints on a Massive Double-Copy and Applications to Massive Gravity}

\author[a]{Laura A. Johnson}
\author[b]{Callum R.~T.~Jones}
\author[b]{Shruti Paranjape}

\affiliation[a]{CERCA, Department of Physics, Case Western Reserve University,\\10900 Euclid Ave, Cleveland, OH 44106}
\affiliation[b]{Leinweber Center for Theoretical Physics, Randall Laboratory of Physics\\ The University of Michigan, Ann Arbor, MI 48109-1040}

\emailAdd{laura.johnson2@case.edu}
\emailAdd{jonescal@umich.edu}
\emailAdd{shrpar@umich.edu}

\abstract{We propose and study a BCJ double-copy of massive particles, showing that it is equivalent to a KLT formula with a kernel given by the inverse of a matrix of massive bi-adjoint scalar amplitudes. For models with a uniform non-zero mass spectrum we demonstrate that the resulting double-copy factors on physical poles and that up to at least 5-particle scattering, color-kinematics satisfying numerators always exist. For the scattering of 5 or more particles, the procedure generically introduces spurious singularities that must be cancelled by imposing additional constraints. When massive particles are present, color-kinematics duality \textit{is not enough} to guarantee a physical double-copy. As an example, we apply the formalism to massive Yang-Mills and show that up to 4-particle scattering the double-copy construction generates physical amplitudes of a model of dRGT massive gravity coupled to a dilaton and a two-form with dilaton parity violating couplings. We show that the spurious singularities in the 5-particle double-copy do not cancel in this example, and the construction fails to generate physically sensible amplitudes. We conjecture sufficient constraints on the mass spectrum, which in addition to massive BCJ relations, guarantee the absence of spurious singularities. }

\begin{document} 
 \begin{flushright}
{\tt LCTP-20-08} \\
\end{flushright}
\maketitle
\flushbottom


\section{Introduction}
\label{sec:intro}

The essence of the \textit{double-copy} is the existence (or conjectured existence) of a map from the physical observables $\mathcal{O}$ of a pair of models $A$ and $B$, each with a non-Abelian internal symmetry structure, to physical observables in some other model $A\otimes B$, without such a symmetry
\begin{equation}
  \mathcal{O}_A \times \mathcal{O}_B \mapsto \mathcal{O}_{A\otimes B}.
\end{equation} 
The original and best-studied example of such a map is given by the construction of Kawai-Lewellen-Tye (KLT), relating tree-level open and closed string scattering amplitudes \cite{Kawai:1985xq}, and the associated field theory limit ($\alpha' \rightarrow 0$) relating Yang-Mills and Einstein gravity. For example at 4-point
\begin{equation}
\label{eq:masslessKLT}
  \mathcal{M}_4^{\text{Grav}}\left(1,2,3,4\right) = -s_{14}\mathcal{A}^{\text{YM}}_4\left[1,2,3,4\right]\mathcal{A}^{\text{YM}}_4\left[1,3,2,4\right].
\end{equation}
The double-copy has subsequently been extended to a surprisingly eclectic class of field theory models including non-linear sigma models and D-brane worldvolume EFTs \cite{Cachazo:2014xea}, generalized to loop-level \cite{Bern:2010ue} and even extended to classes of classical solutions \cite{Monteiro:2014cda}. See \cite{Bern:2019prr} and references therein for a comprehensive review of recent developments. More than a theoretical curiosity, there are often significant practical advantages to making use of such a map whenever it is available. Recent use of a generalized double-copy construction for Feynman integrands in $\mathcal{N}=8$ supergravity allowed the first explicit calculation of 4-point, 5-loop scattering amplitudes \cite{Bern:2018jmv}, a feat that is practically impossible to replicate by other, presently available means. 

It is therefore a timely and relevant theoretical problem to understand the potential scope for generalizing the double-copy, and demarcating the boundary between those models which admit a double-copy structure and those which do not. In this paper we will be concerned with the problem of generalizing the field theory double-copy relation for tree-level scattering amplitudes to models with generic massive spectra. Our central result is the demonstration that when massive particles are present, \textit{color-kinematics duality is not enough} to guarantee a physically well-defined double-copy. We present in detail an explicit example, massive Yang-Mills, for which color-kinematics duality satisfying numerators exist (up to at least $n=5$), but for which the BCJ double-copy prescription generates expressions with non-physical spurious singularities.

The well-known construction of Bern, Carrasco and Johansson (BCJ) \cite{Bern:2008qj} begins by organizing tree-level scattering amplitudes in pure Yang-Mills as a sum over trivalent graphs\footnote{We will use the following convention $c_{12}=f^{a_1a_2 b}f^{a_3 a_4 b}$, $c_{13}=f^{a_1a_3 b}f^{a_4 a_2 b}$ and $c_{14}=f^{a_1a_4 b}f^{a_2 a_3 b}$. We also use Mandelstam invariants with all outgoing momenta, i.e. $s_{ij}=(p_i+p_j)^2.$}
\begin{equation}
\label{BCJform4}
  \mathcal{A}_4\left(1^{a_1},2^{a_2},3^{a_3},4^{a_4}\right) = \frac{c_{12} n_{12}}{s_{12}}+\frac{c_{13}n_{13}}{s_{13}}+\frac{c_{14}n_{14}}{s_{14}}.
\end{equation}
This form of the amplitude reveals the remarkable, hidden property of \textit{color-kinematics duality}, the BCJ numerators satisfy a sum rule
\begin{equation}
\label{kinjac}
  n_{12}+n_{13}+n_{14}=0,
\end{equation}
mirroring the Jacobi relation of the color factors 
\begin{equation}
\label{cjac}
    c_{12}+c_{13}+c_{14}=0.
\end{equation}
Perhaps even more remarkably, making the replacement $c_i\rightarrow n_i$ gives an expression 
\begin{equation}
\label{BCJoutput}
  \mathcal{M}_4\left(1,2,3,4\right) = \frac{n_{12}^2}{s_{12}}+\frac{n_{13}^2}{s_{13}}+\frac{n_{14}^2}{s_{14}},
\end{equation}
which has all of the formal properties (gauge invariance, unitarity, locality) of a scattering amplitude in a model of Einstein gravity coupled to a massless dilaton and Kalb-Ramond two-form. The fundamental result of BCJ was to prove this \textit{BCJ double-copy} and the property of color-kinematics duality persist at all multiplicity.

An important point to emphasize is that even though the BCJ numerators are non-unique, since the amplitude (\ref{BCJform4}) is unchanged by a \textit{generalized gauge transformation} 
\begin{equation}
\label{gengauge}
  n_{12}\rightarrow n_{12} + s_{12}\Delta, \hspace{5mm}  n_{13}\rightarrow n_{13} + s_{13}\Delta, \hspace{5mm}  n_{14}\rightarrow n_{14} + s_{14}\Delta,
\end{equation}
where $\Delta$ is an arbitrary function, this freedom is \textit{insufficient in general} to find a BCJ representation satisfying color-kinematics duality. Due to the kinematic identity $s_{12}+s_{13}+s_{14}=0$, the value of the Jacobi sum of numerators $n_{12}+n_{13}+n_{14}$ is also invariant. If the kinematic Jacobi relation is violated in one generalized gauge, it will be violated in every generalized gauge. The existence of color-kinematics duality satisfying numerators is then a very special property which is obtained only in Yang-Mills and a handful of other models \cite{Bern:2019prr}.

While beautifully simple, it is not at all obvious that the expression (\ref{BCJoutput}) resulting from the BCJ construction is a physical scattering amplitude. In particular, this construction fails to manifest \textit{locality} in the form of the absence of spurious, non-propagator-like, singularities and the factorization of amplitudes on propagator-like, physical singularities. For $n\geq 5$ the generalized gauge functions needed to bring the local form of BCJ numerators generated by Feynman rules, to a color-kinematics duality satisfying representation, can in principle be arbitrarily complicated. The simplest way to show the absence of spurious singularities in the BCJ construction is to prove that it is equivalent to the KLT construction which, after making a convenient choice of basis, can be written in a form that manifests the absence of spurious singularities \cite{Bern:2008qj}. 

The BCJ construction has a natural extension to models containing massive states, for which various special cases have been considered previously \cite{Naculich:2014naa,Johansson:2015oia,Johansson:2019dnu,Chiodaroli:2015rdg,Chiodaroli:2017ehv,Chiodaroli:2018dbu,bautista2019double,bautista2019scattering,Neill_2013,Bjerrum_Bohr_2019,Bern_2019}. To our knowledge, no completely general description of a massive BCJ double-copy, and the associated constraints, has been given. In particular, the case of double-copying amplitudes in theories with no massless particles has not been studied before. This paper is a first step towards such a description. The direct analogue of the BCJ form of the amplitude for models with a uniform, non-zero mass spectrum is\footnote{Throughout this paper we will use the mostly-plus metric convention $\eta_{\mu\nu} = \text{diag}\left(-1,+1,+1,+1\right)$.} 
\begin{equation}
\label{mBCJform4}
  \mathcal{A}^{m\neq 0}_4\left(1^{a_1},2^{a_2},3^{a_3},4^{a_4}\right) = \frac{c_{12} n_{12}}{s_{12}+m^2}+\frac{c_{13}n_{13}}{s_{13}+m^2}+\frac{c_{14}n_{14}}{s_{14}+m^2}.
\end{equation}
To construct the \textit{massive double-copy} of such a model, we will follow closely the discussion above, and try to construct numerators which satisfy the kinematic Jacobi relation $n_{12}+n_{13}+n_{14}=0$. If we succeed, we make the replacement $c_i\rightarrow n_i$ and construct the would-be massive double-copy
\begin{equation}
\label{mBCJoutput}
  \mathcal{M}^{m\neq 0}_4\left(1,2,3,4\right) = \frac{n_{12}^2}{s_{12}+m^2}+\frac{n_{13}^2}{s_{13}+m^2}+\frac{n_{14}^2}{s_{14}+m^2}.
\end{equation} 
The central problem in this paper will be to understand the conditions under which expressions such as (\ref{mBCJoutput}) and its natural generalization to higher multiplicity, define physical scattering amplitudes. At this point we make a simple observation, the massive analogue of a generalized gauge transformation 
\begin{equation}
\label{eq:ggt}
  n_{12}\rightarrow n_{12} + (s_{12}+m^2)\Delta, \hspace{5mm}  n_{13}\rightarrow n_{13} + (s_{13}+m^2)\Delta, \hspace{5mm}  n_{14}\rightarrow n_{14} + (s_{14}+m^2)\Delta,
\end{equation}
 which leaves (\ref{mBCJform4}) invariant, does \textit{not} leave the sum of numerators invariant, rather
\begin{equation}
  n_{12}+n_{13}+n_{14} \rightarrow n_{12}+n_{13}+n_{14} -m^2 \Delta.
\end{equation}
It follows that, when $m\neq 0$, we can \textit{always} find a generalized gauge that realizes color-kinematics duality! Since this argument relied only on knowledge of the spectrum, it applies to all models with uniform non-zero mass spectra, independent of the details of the interactions. Contrary to the $m=0$ case where color-kinematics duality was a special property only found in a handful of models, for models with generic massive spectra it is no constraint at all. As we will see, this situation is indeed too good to be true. By rewriting the resulting would-be double-copy in a KLT-like form with a kernel given by the inverse of a matrix of massive bi-adjoint scalar amplitudes \cite{Cachazo:2013iea,Mizera:2016jhj}, we find that the double-copy generically introduces non-local, spurious singularities for $n>4$, and fails to reduce to the standard double-copy as $m\rightarrow 0$.

As a theoretical laboratory for making explicit calculations, we consider the physically well-motivated example of a model of massive Yang-Mills. As we explain in detail in Section \ref{sec:mYMhighE}, by considering the reduction to the familiar massless double-copy in the high-energy or Goldstone boson equivalence limit, there is a plausible expectation that massive Yang-Mills double copies to a model of de Rham-Gabadadze-Tolley or \textit{dRGT massive gravity} \cite{deRham:2010kj} coupled to a massive dilaton and a massive two-form. The primary conclusion of this example is that no miraculous cancellation of the spurious singularities takes place, and the proposed massive double-copy fails to generate physical scattering amplitudes for $n>4$. We will conclude by revisiting the logic of the above argument in Section \ref{sec:diffmass} and demonstrate that if, in addition to color-kinematics duality, the spectrum of masses satisfies certain constraints, then a local massive double-copy does indeed exist.

An outline of this paper is as follows. In Section \ref{sec:massKLT} we show that the massive BCJ double-copy can be equivalently formulated as a KLT-like product with a kernel given by the inverse of a $(n-2)!\times (n-2)!$ matrix of massive bi-adjoint scalar amplitudes. Here we use the Del Duca-Dixon-Maltoni (DDM) basis \cite{DelDuca:1999rs}. The KLT product is shown to factor into lower-point amplitudes on physical poles at all multiplicities, but generically does not smoothly reduce to the  massless KLT product as $m\rightarrow 0$. Additionally, for $n\geq 5$ the massive kernel contains spurious singularities that cannot be associated with a factorization channel for any physical state. In Section \ref{sec:mYM2} we present our primary explicit example, the double-copy of a mass-deformed version of Yang-Mills. At $n=3$ and $n=4$, the double-copy gives physically sensible results that can be interpreted as scattering amplitudes of a model of dRGT massive gravity coupled to a dilaton and two-form with a $\Lambda_3$ cutoff scale. At $n=5$, we numerically evaluate the residue on the spurious singularities, and confirm that they are non-zero, demonstrating that the BCJ double-copy does not produce a physical scattering amplitude. In Section \ref{sec:diffmass} we consider models with a general spectrum of masses. We show that if a certain condition is imposed on the spectrum, then the rank of the bi-adjoint scalar matrix is reduced and implies massive versions of the fundamental BCJ relations. It is shown that if the rank is reduced to $(n-3)!$ then the massive double-copy takes a manifestly local form which reduces smoothly to the massless double-copy. In Section \ref{sec:conclusions} we conclude and describe important future directions.

While this work was in its final stages, the preprint \cite{Momeni:2020vvr} by Momeni, Rumbutis and Tolley appeared with some overlapping results at 4-point level.

\section{Massive KLT Formula}
\label{sec:massKLT}

Models with on-shell $U(N)$ symmetry, with asymptotic states in the adjoint representation, admit a convenient decomposition of tree-amplitudes into \textit{single trace} or \textit{color-ordered partial} amplitudes of the form
\begin{equation}
\label{partial}
    \mathcal{A}_n\left(1^{a_1},...,n^{a_n}\right) = \sum_{\sigma \in S_{n-1}}\text{Tr}\left[T^{a_1}T^{a_{\sigma(2)}}...T^{a_{\sigma(n)}}\right]\mathcal{A}_n\left[1\;\sigma(2)...\sigma(n)\right].
\end{equation}
For such models, the BCJ double-copy is formally equivalent to the field theory limit of the KLT relations
\begin{align}
\label{eq:KLT}
\mathcal{A}^{A\otimes B}_n(1,2,\cdots,n)&=\sum_{\alpha,\beta}\mathcal{A}^\text{A}_n[\alpha]\, S[\alpha|\beta]\mathcal{A}^\text{B}_n[\beta],
\end{align}
where the sum is taken over (possibly distinct) \textit{BCJ bases} of size $(n-3)!$, and $S$ is a function of Mandelstam invariants called the KLT kernel. An explicit all-multiplicity expression for the massless KLT kernel, in a particularly convenient choice of BCJ basis, is given in eq. (2.51) of \cite{Bern:2019prr}. 

In Section \ref{sec:intro}, we saw a proposed extension of the BCJ double-copy to models with uniform, non-zero mass spectra. In this section, by following the same well-known formal manipulations as the massless case, we will derive an equivalent \textit{massive KLT formula}. Let us begin by illustrating the argument explicitly at 4-point. We consider two, possibly distinct models $A$ and $B$ with on-shell $U(N)$ symmetry and uniform mass spectra, which admit a BCJ representation (\ref{mBCJform4}). If we can find a generalized gauge for which the numerators satisfy the kinematic Jacobi relations
\begin{equation}
\label{kinjacAB}
    n_{12}^A+n_{13}^A+n_{14}^A = 0, \hspace{5mm} n_{12}^B+n_{13}^B+n_{14}^B = 0,
\end{equation}
then the proposed massive BCJ double-copy takes the form
\begin{align}
\label{eq:BCJmassive}
\mathcal{A}^{A\otimes B}_4(1,2,3,4)&=\frac{n^A_{12} n^B_{12}}{s_{12}+m^2}+\frac{n^A_{13}n^B_{13}}{s_{13}+m^2}+\frac{n^A_{14}n^B_{14}}{s_{14}+m^2}.
\end{align}
As mentioned in Section \ref{sec:intro}, for a model with this mass spectrum a set of kinematic Jacobi-satisfying numerators can \textit{always} be found. Consider for example a set of numerators, perhaps generated from Feynman rules, that do not satisfy the kinematic Jacobi relations,
\begin{align}
n_{12}+n_{13}+n_{14} = \mathcal{E}\ne 0.
\end{align}
We can always perform the following generalized gauge transformation,
\begin{align}
\label{gaugeE}
n_{12}&\to n_{12}+\frac1{m^2}(s_{12}+m^2)\mathcal{E} \nonumber\\
n_{13}&\to n_{13}+\frac1{m^2}(s_{13}+m^2)\mathcal{E} \nonumber\\
n_{14}&\to n_{14}+\frac1{m^2}(s_{14}+m^2)\mathcal{E},
\end{align}
to generate a set of numerators satisfying (\ref{kinjacAB}).

It is useful to rephrase this argument in a way that makes natural the extension to all-multiplicity. We begin with the important observation that the number of linearly independent numerators, after imposing kinematic Jacobi identities, is $(n-2)!$ which is exactly the number of amplitudes in a DDM basis \cite{DelDuca:1999rs}. By decomposing the color-factors in the BCJ representation into single-trace components, the color-ordered partial amplitudes (\ref{kinjacAB}) can be expressed as linear functions of the numerators, for example
\begin{align}
\mathcal{A}_4[1234]&=\frac{n_{12}}{s_{12}+m^2}-\frac{n_{14}}{s_{14}+m^2}.
\end{align} 
For a model admitting both a BCJ representation (\ref{BCJform4}) \textit{and} a single-trace decomposition (\ref{partial}), the $(n-1)!$ non-cyclically related partial amplitudes cannot be linearly independent. Indeed, by reducing to a DDM basis of color-structures \cite{DelDuca:1999rs}, the number of independent partial amplitudes is found to be $(n-2)!$ with the non-trivial linear relations among the original $(n-1)!$ basis given by the so-called \textit{Kleiss-Kuijf (KK) relations} \cite{Kleiss:1988ne}. This counting leads us to the conclusion that there are an equal number of independent partial amplitudes and kinematic numerators, which are linearly related by equations of the form
\begin{align}
\label{APn}
\begin{pmatrix}
\mathcal{A}_4[1234]\\
\mathcal{A}_4[1324]
\end{pmatrix}=&\begin{pmatrix}
\frac{1}{s_{12}+m^2}+\frac{1}{s_{14}+m^2}  && \frac{1}{s_{14}+m^2} \\
-\frac{1}{s_{14}+m^2}  && -\frac{1}{s_{13}+m^2} -\frac{1}{s_{14}+m^2} 
\end{pmatrix}\begin{pmatrix}
n_{12} \\ n_{13}
\end{pmatrix}.
\end{align}
If $m\neq 0$, then the propagator matrix has full-rank and so we can \textit{solve} for the kinematic Jacobi-satisfying numerators
\begin{align}
\label{eq:numsol}
\begin{pmatrix}
n_{12} \\ n_{13}
\end{pmatrix}=&\begin{pmatrix}
\frac{1}{s_{12}+m^2}+\frac{1}{s_{14}+m^2}  && \frac{1}{s_{14}+m^2} \\
-\frac{1}{s_{14}+m^2}  && -\frac{1}{s_{13}+m^2} -\frac{1}{s_{14}+m^2} 
\end{pmatrix}^{-1}\begin{pmatrix}
\mathcal{A}_4[1234] \\ \mathcal{A}_4[1324]
\end{pmatrix}.
\end{align}
When $m = 0$ however, the propagator matrix has rank 1, and no such inversion is possible. In this case, the massless propagator matrix has a null-vector, and so we can make the replacement
\begin{equation}
    \begin{pmatrix}
        n_{12} \\
        n_{13}
    \end{pmatrix}
    \rightarrow
     \begin{pmatrix}
        \hat{n}_{12} \\
        \hat{n}_{13}
    \end{pmatrix}
    =
    \begin{pmatrix}
        n_{12} \\
        n_{13}
    \end{pmatrix}
    +\Delta 
    \begin{pmatrix}
        s_{12} \\
        s_{13}
    \end{pmatrix},
\end{equation}
for any function $\Delta$. We recognize this is as the statement that generalized gauge invariance (\ref{gengauge}) preserves the kinematic Jacobi relations (\ref{kinjacAB}). We can use this residual freedom to impose the gauge-fixing conditions $\hat{n}_{13}=0$, and solve for $\hat{n}_{12}$. From (\ref{APn}) with $m=0$, we have two different expressions for $\hat{n}_{12}$ which must be equal, leading to the so-called \textit{fundamental BCJ identity}
\begin{equation}
    s_{12}\mathcal{A}_4\left[1234\right] = s_{13}\mathcal{A}_4[1324].
\end{equation}
We can run this argument in both directions, reaching the well-known conclusion that the fundamental BCJ relations are \textit{necessary} and \textit{sufficient} conditions for the existence of color-kinematics duality satisfying BCJ numerators. 

This analysis generalizes naturally to $n$-point. For a model with uniform mass spectrum and $m\neq 0$ there is a linear relation of the form (\ref{APn}) relating the $(n-2)!$ DDM bases of partial amplitudes and the kinematic numerators with an $(n-2)!\times(n-2)!$ propagator matrix. We believe that this matrix is always of full-rank, but do not have a proof of this fact. An explicit expression for the $6\times 6$ massive propagator matrix at $n=5$ is given in Appendix \ref{app:5ptmatrix}, from which the rank can be verified to be 6. 

If the matrix is full-rank, then we can solve for a \textit{unique} set of color-kinematics duality satisfying numerators by inverting the linear relation. Furthermore, since a full-rank propagator matrix has no null vectors, there are no additional BCJ-like relations among the partial amplitudes and hence no constraints on which models can admit a massive BCJ double-copy. As we will discuss further in Section \ref{sec:diffmass}, these conclusions may be modified in models with a more complicated spectrum of masses. 

Once we have solved for the numerators, it is straightforward to rewrite the BCJ double-copy in KLT form (\ref{eq:KLT}). We first rewrite the BCJ double-copy \eqref{eq:BCJmassive} in matrix form
\begin{align}
\mathcal{A}^{A\otimes B}_4(1,2,3,4)&=\begin{pmatrix}
n^A_{12} && n^A_{13}
\end{pmatrix}\begin{pmatrix}
\frac{1}{s_{12}+m^2}+\frac{1}{s_{14}+m^2}  && -\frac{1}{s_{14}+m^2} \\
-\frac{1}{s_{14}+m^2}  && \frac{1}{s_{13}+m^2} +\frac{1}{s_{14}+m^2}
\end{pmatrix} 
\begin{pmatrix}
n^B_{12} \\ n^B_{13}
\end{pmatrix},
\end{align}
where we have already imposed the kinematic Jacobi identities (\ref{kinjacAB}). Combining this with our solution for the numerators \eqref{eq:numsol} gives
\begin{align}
\label{eq:KLTBCJ4}
\small
&\mathcal{A}^{A\otimes B}_4(1,2,3,4)\nonumber\\
&\hspace{0.5cm}=\begin{pmatrix}
\mathcal{A}^A_4[1234] &&  \mathcal{A}^A_4[1324]
\end{pmatrix}\begin{pmatrix}
\frac{1}{s_{12}+m^2}+\frac{1}{s_{14}+m^2}  && \hspace{-0cm}-\frac{1}{s_{14}+m^2} \\
-\frac{1}{s_{14}+m^2}  && \hspace{-0cm}\frac{1}{s_{13}+m^2} +\frac{1}{s_{14}+m^2}
\end{pmatrix}^{-1}\hspace{-0.25cm}
\begin{pmatrix}
\mathcal{A}^B_4[1234] \\ \mathcal{A}^B_4[1324]
\end{pmatrix},
\end{align}
which is of KLT form, with the matrix in the middle acting as a \textit{massive KLT kernel}.

A similar calculation can be performed at 5-point, both to calculate the 6 numerators from a DDM basis of 6 amplitudes and to calculate the KLT kernel. The details of this calculation are presented in Appendix \ref{app:5ptmatrix}.

\subsection{Massive KLT Kernel}
\label{sec:mKLT} 
Somewhat recently the massless KLT kernel was understood to be the inverse of a matrix of bi-adjoint scalar amplitudes \cite{Cachazo:2013iea,Mizera:2016jhj}. Previously, we saw how to construct a KLT kernel from a massive BCJ double-copy. In this section we recognize this kernel as the inverse of a matrix of \textit{massive} bi-adjoint scalar amplitudes, giving us a general prescription for an $n$-point massive KLT formula.

We begin by reviewing the construction in the massless case. The scattering amplitudes of the following $U(N)\times U(\tilde{N})$ invariant model of massless scalars transforming in the bi-adjoint representation 
\begin{equation}
    \mathcal{L} = -\frac{1}{2}\left(\partial_\mu \phi^{aa'}\right)^2 - gf^{abc}\tilde{f}^{a'b'c'}\phi^{aa'}\phi^{bb'}\phi^{cc'},
\end{equation}
admit a \textit{double} color-ordering
\begin{equation}
    \mathcal{A}^{\phi^3}_n\left(1^{a_1 a'_1},...,n^{a_n a'_n}\right) = \sum_{\alpha,\beta \in S_{n-1}}\text{Tr}\Big[T^{a_1}T^{a_{\alpha(2)}}...T^{a_{\alpha(n)}}\Big]\text{Tr}\left[\tilde{T}^{a'_1}\tilde{T}^{a'_{\beta(2)}}...\tilde{T}^{a'_{\beta(n)}}\right]\mathcal{A}_n^{\phi^3}\left[\alpha|\beta\right].
\end{equation}
The partial amplitudes $\mathcal{A}_n^{\phi^3}\left[\alpha|\beta\right]$ are indexed by two orderings and can be constructed efficiently via a simple diagrammatic procedure \cite{Mizera:2016jhj}. Regarding $\mathcal{A}^{\phi^3}[\alpha|\beta]$ as an $(n-1)!\times (n-1)!$ matrix, it can be shown to have rank $(n-3)!$ \cite{Cachazo:2013iea}. The null vectors correspond to separate  row and column KK and BCJ relations. For example at 4-point
\begin{align}
s_{12}\mathcal{A}^{\phi^3}_4[1234|1234]=s_{13}\mathcal{A}^{\phi^3}_4[1324|1234].
\end{align}
The central result of \cite{Cachazo:2013iea} was to prove that a BCJ-independent $(n-3)!\times (n-3)!$ sub-matrix has full-rank, and moreover has an inverse which is precisely equal to the KLT kernel in the given BCJ basis
\begin{equation}
    S[\alpha|\beta] = \left(\mathcal{A}_n^{\phi^3}\right)^{-1}\left[\alpha|\beta\right].
\end{equation}
The massless KLT formula (\ref{eq:KLT}) can then be equivalently formulated as
\begin{align}
\label{eq:KLTbiadjoint}
\mathcal{A}^{A\otimes B}_n(1,2,\cdots,n)&=\sum_{\alpha,\beta}\mathcal{A}^A_n[\alpha]\,\, \left(\mathcal{A}^{\phi^3}\right)^{-1}[\alpha|\beta]\mathcal{A}^B_n[\beta].
\end{align}
Let us now investigate what happens in \textit{massive bi-adjoint scalar theory}
\begin{equation}
    \mathcal{L} = -\frac{1}{2}\left(\partial_\mu \phi^{aa'}\right)^2-\frac{1}{2}m^2 \phi^{aa'}\phi^{aa'} - gf^{abc}\tilde{f}^{a'b'c'}\phi^{aa'}\phi^{bb'}\phi^{cc'}.
\end{equation}
Amplitudes in the massive theory are constructed using the same diagrammatic rules used for the massless theory \cite{Mizera:2016jhj}, but with the massless propagators replaced with their massive counterparts. For example,
\begin{align}
\mathcal{A}^{\phi^3}_4[1234|1234]&=\frac{1}{s_{12}+m^2}+\frac{1}{s_{14}+m^2},\\
\mathcal{A}^{\phi^3}_4[1234|1324]&=-\frac{1}{s_{14}+m^2}\,.
\end{align}

The 5-point matrix of bi-adjoint scalar amplitudes can be found in Appendix \ref{app:5ptmatrix}. The primary difference between the massless and massive bi-adjoint scalar amplitudes is in the number of independent color-orderings. In the massive theory, DDM orderings are independent and the $(n-2)!\times (n-2)!$ matrix of bi-adjoint scalar amplitudes has full-rank. Since this matrix is invertible, there is a natural conjecture for a massive KLT formula. At 4-point this takes the explicit form
\begin{align}
\label{eq:KLT4pt}
&\hspace{-4mm}\mathcal{A}^{A\otimes B}_4 (1,2,3,4) \nonumber\\
=&\begin{pmatrix}
\mathcal{A}_4^A[1234] && \mathcal{A}_4^A[1324]
\end{pmatrix} \begin{pmatrix}
\mathcal{A}^{\phi^3}_4[1234|1234] && \mathcal{A}^{\phi^3}_4[1234|1324]\\
\mathcal{A}^{\phi^3}_4[1234|1324] && \mathcal{A}^{\phi^3}_4[1324|1324]
\end{pmatrix}^{-1}\begin{pmatrix}
\mathcal{A}_4^B[1234] \\ \mathcal{A}_4^B[1324]
\end{pmatrix}\,\nonumber\\
=&\;\frac{1}{m^2}\mathcal{A}_4^A[1234] \left(m^2+s_{12}\right) \left(\mathcal{A}_4^B[1234] \left(2 m^2+s_{12}\right)-\mathcal{A}_4^B[1324]\left(m^2+s_{13}\right)\right)\nonumber\\
&+\frac{1}{m^2}\mathcal{A}_4^A[1324] \left(m^2+s_{13}\right)\left(-\mathcal{A}_4^B[1234]\left(m^2+s_{12}\right)+\mathcal{A}_4^B[1324] \left(2m^2+s_{13}\right)\right).
\end{align}
This formula is exactly the one we arrived at from the massive BCJ double-copy in \eqref{eq:KLTBCJ4}. An interesting aspect of the massive KLT formula is its massless limit in which,
\begin{align}
\mathcal{A}^{A\otimes B}_4 (1,2,3,4)=&\frac{1}{m^2}\left( s_{12}\mathcal{A}^A_4\left[1234\right] - s_{13}\mathcal{A}^A_4[1324]\right)\left( s_{12}\mathcal{A}^B_4\left[1234\right] - s_{13}\mathcal{A}^B_4[1324]\right)\nonumber\\
&+\left(3 s_{12}\mathcal{A}^A_4\left[1234\right]\mathcal{A}^B_4\left[1234\right] +3 s_{13}\mathcal{A}^A_4\left[1324\right]\mathcal{A}^B_4\left[1324\right] \right.\nonumber\\
&\left.+s_{14}\left(\mathcal{A}^A_4\left[1234\right]\mathcal{A}^B_4\left[1324\right]+\mathcal{A}^A_4\left[1324\right]\mathcal{A}^B_4\left[1234\right]\right)\right)+\mathcal{O}(m^2)\,.
\end{align}
The coefficient of the leading $\mathcal{O}(m^{-2})$ term is recognizable as a product of fundamental BCJ relations. Thus, we see that the formula admits a smooth massless limit only when the fundamental BCJ relation is satisfied either by theory $A$ or $B$ in their massless limits. For a pair of massive deformations of BCJ-compatible theories, the massive KLT formula then reduces to the familiar massless KLT relation,
\begin{align}
    \mathcal{A}^{A\otimes B}_4 (1,2,3,4)=&-s_{14}\mathcal{A}^A[1234]\mathcal{A}^B[1324]+\mathcal{O}(m^2)\,.
\end{align}

We now proceed to 5-point, where the explicit comparison of KLT and BCJ forms of the double-copy can be repeated using the results of Appendix \ref{app:5ptmatrix} to find that, again, the KLT kernel from the massive BCJ double-copy is precisely the inverse of massive bi-adjoint scalar amplitudes.

For general $n$, we propose the following massive KLT formula
\begin{align}
\label{eq:KLTmassive}
\mathcal{A}^{A \otimes B}_n(1,2,\cdots,n)&=\sum_{\alpha,\beta}\mathcal{A}^A_n[\alpha]\,\, \left(\mathcal{A}_n^{\phi^3}\right)^{-1}[\alpha|\beta]\,\,\mathcal{A}^B_n[\beta],
\end{align}
where $\alpha$ and $\beta$ now range over all $(n-2)!$ DDM color orderings and $\mathcal{A}_n^{\phi^3}[\alpha|\beta]$ is a matrix of amplitudes of massive bi-adjoint scalar theory. We have shown explicitly that up to $n=5$ this agrees with the massive BCJ double-copy and conjecture that they agree at all $n$.

\subsection{Factorization on Physical Poles}
\label{sec:factor}
An essential property of amplitudes in local theories is the presence of simple poles when intermediate momenta go on-shell and factorization of the amplitude into products of lower-point amplitudes in the associated residue. In this section, we will discuss how these properties are ensured in amplitudes generated by our proposed massive KLT formula \eqref{eq:KLTmassive}. We will focus on factorization of \eqref{eq:KLTmassive} on two-particle channels and relegate the discussion of multi-particle channels to Appendix \ref{app:factor}. 

We begin by assuming that theories $A$ and $B$ are local and that their amplitudes factorize correctly on two-particle channels,
\begin{align}
\label{eq:ABfact}
\mathcal{A}_n^A[12,\sigma] &= \frac{\mathcal{A}_3^A[12,-P_{12}]\mathcal{A}_{n-1}^A[P_{12},\sigma]}{s_{12}+m^2} + \mathcal{O}\left((s_{12}+m^2)^{0}\right) \nonumber\\
\mathcal{A}_n^B[12,\sigma] &= \frac{\mathcal{A}_3^B[12,-P_{12}]\mathcal{A}_{n-1}^B[P_{12},\sigma]}{s_{12}+m^2} + \mathcal{O}\left((s_{12}+m^2)^{0}\right) ,
\end{align}
where there is an implicit sum over states on the right-hand-side. 

Next without loss of generality, let us choose to study factorization on the $s_{12}$ pole. We can further assume that we have chosen a DDM basis in which the first $m$ elements have the form $[12\sigma(3,\cdots,n)]$ where $\sigma$ is a permutation, and no other elements have 1 and 2 adjacent\footnote{For example, the basis $[1\sigma(2,\cdots,n-1)n]$ where $\sigma$ runs over all $(n-2)!$ permutations is a DDM basis with $(n-3)!$ elements of the form $[12\sigma(3,\cdots,n-1)n]$. One can then choose an ordering of basis elements such that the assumed property is fulfilled.}. Thus only orderings in the first $m$ rows and columns admit poles in $s_{12}$ and we can resolve our matrix of bi-adjoint scalar amplitudes into blocks,
\begin{equation}
\mathcal{A}_n^{\phi^3}[\alpha|\beta] = 
\begin{pmatrix}
P & Q^\top \\
Q & R
\end{pmatrix},
\end{equation}
where $P$, $Q$ and $R$ are $m\times m$, $(n-m)\times m$ and $(n-m)\times(n-m)$ matrices respectively. Since $s_{12}$ poles are not admitted by the last $(n-m)$ orderings, the $P$ matrix contains all the elements with an $s_{12}$ pole and $Q$ and $R$ do not contain any elements with an $s_{12}$ pole. Locality and unitarity of bi-adjoint scalar theory then demands that elements of $Q$ and $R$ will have zero residue on the $s_{12}$ pole, and a given element of $P$ will have the form
\begin{align}
\label{eq:Ainv}
\mathcal{A}_n^{\phi^3}[12,\sigma(3,\cdots,n)|12,\sigma'(3,\cdots,n)] =& \frac{\mathcal{A}_{n-1}^{\phi^3}[P_{12},\sigma(3,\cdots,n)|P_{12},\sigma'(3,\cdots,n)]}{s_{12}+m^2} \nonumber\\
&\hspace{4cm}+ \mathcal{O}\left((s_{12}+m^2)^{0}\right),
\end{align}
near the pole. We will now assume that the orderings $[P_{12},\sigma(3,\cdots,n)]$ form a DDM basis for $n-1$ particles $\{P_{12},3,4,...,n\}$\footnote{Returning to the example DDM basis $[1\sigma(2,\cdots,n-1)n]$, we see that this condition is satisfied, i.e. $[P_{12}\sigma(3,\cdots,n-1)n]$ forms a DDM basis.}. Thus the blocks are characterized by their behavior as they approach the $s_{12}$ pole,
\begin{equation}
P = \mathcal{O}\left((s_{12}+m^2)^{-1}\right), \hspace{5mm} Q = \mathcal{O}\left((s_{12}+m^2)^{0}\right), \hspace{5mm} R = \mathcal{O}\left((s_{12}+m^2)^{0}\right).
\end{equation}

Various useful corollaries can be drawn. For example,
\begin{equation}
P^{-1} = \mathcal{O}\left((s_{12}+m^2)^{1}\right), \hspace{5mm}  R^{-1} = \mathcal{O}\left((s_{12}+m^2)^{0}\right).
\end{equation}
In fact, \eqref{eq:Ainv} allows us to be more specific, for $P^{-1}$,
\begin{equation}
P^{-1}[12,\sigma|12,\sigma']  = (s_{12}+m^2) \left(\mathcal{A}_{n-1}^{\phi^3}\right)^{-1} [P_{12},\sigma|P_{12},\sigma'] + \mathcal{O}\left((s_{12}+m^2)^{2}\right),
\end{equation}
where we will use the shorthand $\sigma=\sigma(3,\cdots,n)$ and $\sigma'=\sigma'(3,\cdots,n)$ for the rest of the section. Finally, using the geometric series formula for matrices, we get
\begin{equation}
(1-P^{-1}Q^\top R^{-1}Q)^{-1} = 1 + \mathcal{O}\left((s_{12}+m^2)^{1}\right).
\end{equation}
These properties, along with the blockwise inversion formula
\begin{equation}
\label{eq:invblock}
\footnotesize
\left(\mathcal{A}_n^{\phi^3}\right)^{-1}[\alpha|\beta] = 
\begin{pmatrix}
P^{-1}(1-P^{-1}Q^\top R^{-1}Q)^{-1} & -P^{-1}(1-P^{-1}Q^\top R^{-1}Q)^{-1}Q^\top R^{-1} \\
-R^{-1}QP^{-1}(1-P^{-1}Q^\top R^{-1}Q)^{-1} & R^{-1}+R^{-1}QP^{-1}(1-P^{-1}Q^\top R^{-1}Q)^{-1}Q^\top R^{-1}
\end{pmatrix},
\end{equation}
gives 
\begin{equation}
\footnotesize
\begin{pmatrix}
P & Q^\top \\
Q & R
\end{pmatrix}^{-1} = 
\begin{pmatrix}
(s_{12}+m^2) \left(\mathcal{A}_{n-1}^{\phi^3}\right)^{-1} [P_{12},\sigma|P_{12},\sigma']  & 0\\
0 & 0
\end{pmatrix}
+
\begin{pmatrix}
\mathcal{O}\left((s_{12}+m^2)^{2}\right) &\mathcal{O}\left((s_{12}+m^2)^{1}\right) \\
\mathcal{O}\left((s_{12}+m^2)^{1}\right) & \mathcal{O}\left((s_{12}+m^2)^{0}\right)
\end{pmatrix}.
\end{equation}
It is straightforward to see that only the elements in the top left block will multiply amplitudes $\mathcal{A}_3^A[12\sigma]$ and $\mathcal{A}_3^B[12\sigma']$ and hence only these could develop a pole at $s_{12}$. Thus the suppressed terms on the right-hand-side will not contribute on the factorization channel. 

So in a neighborhood of the $s_{12}$ pole,
\begin{align}
\mathcal{A}_n^{A\otimes B} &= \sum_{\alpha,\beta} \mathcal{A}_n^A[\alpha] \left(\mathcal{A}_n^{\phi^3}\right)^{-1}[\alpha|\beta] \mathcal{A}^B_n[\beta] \nonumber\\
&= \sum_{\sigma,\sigma'} \frac{\mathcal{A}_3^A[12,-P_{12}]\mathcal{A}_{n-1}^A[P_{12},\sigma]\left(\mathcal{A}_{n-1}^{\phi^3}\right)^{-1}\hspace{-1mm}[P_{12},\sigma|P_{12},\sigma'] \mathcal{A}_3^B[12,-P_{12}]\mathcal{A}_{n-1}^B[P_{12},\sigma']}{s_{12}+m^2} \nonumber\\
&\hspace{10mm} +  \mathcal{O}\left((s_{12}+m^2)^{0}\right) \nonumber\\
\hspace{1cm}&=\mathcal{A}_3^A[12,-P_{12}]\mathcal{A}_3^B[12,-P_{12}] \sum_{\sigma,\sigma'} \frac{\mathcal{A}_{n-1}^A[P_{12},\sigma]\left(\mathcal{A}_{n-1}^{\phi^3}\right)^{-1}\hspace{-1mm} [P_{12},\sigma|P_{12},\sigma'] \mathcal{A}_{n-1}^B[P_{12},\sigma']}{s_{12}+m^2} \nonumber\\
&\hspace{10mm} +  \mathcal{O}\left((s_{12}+m^2)^{0}\right) \nonumber\\ \nonumber\\
&= \frac{\mathcal{A}_3^{A\otimes B}(1,2,-P_{12}) \mathcal{A}_{n-1}^{A\otimes B}(P_{12},3,...,n)}{s_{12}+m^2} +  \mathcal{O}\left((s_{12}+m^2)^{0}\right),
\end{align}

where we have used the fact that for $n=3$, the formula \eqref{eq:KLTmassive} takes the simple form,
\begin{equation}
\mathcal{A}^{A\otimes B}_3(1,2,3) = \mathcal{A}_3^A[123]\mathcal{A}_3^B[123].
\end{equation}

Thus, on a two-particle channel, an $n$-point amplitude generated by the massive KLT formula factorizes into lower-point amplitudes also generated by \eqref{eq:KLTmassive}, i.e. these amplitudes factorize into the correct lower-point amplitudes. Since we chose to study the $s_{12}$ pole without loss of generality, this argument demonstrates factorization on all two-particle singularities.

\subsection{Spurious Poles}
\label{sec:spurious}
We have seen so far that our proposed massive KLT formula has various nice properties. It does not require BCJ-type constraints in order to get a consistent answer, allowing any theory to be ``double-copied''. In addition, it manifestly creates a set of amplitudes that factorize into lower-point amplitudes on physical poles. These properties might suggest that \eqref{eq:KLTmassive} can double-copy any massive theory into a different local theory, but this is not the case. While we saw in the previous section that all necessary physical poles are present in \eqref{eq:KLTmassive}, we did not check that these are the only poles, i.e. we did not test the presence of spurious/non-physical poles in amplitudes resulting from \eqref{eq:KLTmassive}. 

In general the inverse of a matrix $A^{-1}$ equals the matrix of cofactors times $1/\text{det}A$. Cofactors are sums of products of elements of $A$ and hence no new poles can be generated in the matrix of cofactors. Thus, all new poles must be a result of zeros of det$A$. Applying this to \eqref{eq:KLTmassive}, we find that physical poles will appear in the matrix of cofactors, while spurious poles could occur due to zeros of the determinant of $\mathcal{A}^{\phi^3}_n\left[\alpha|\beta\right]$. 

Let us first understand how this is taken care of in the massless case. Here the BCJ relations restrict us to a subset of the DDM basis. As a result, not all physical poles are present in $\mathcal{A}^{\phi^3}_n\left[\alpha|\beta\right]$. Thus some physical poles must appear as zeros of the determinant of $\mathcal{A}^{\phi^3}_n\left[\alpha|\beta\right]$, while others will appear in the matrix of cofactors. For example at 4-point we have
\begin{align}
\mathcal{A}_4^{\phi^3}[1234|1234]&=-\frac{s_{13}}{s_{12}s_{14}}\\
\Rightarrow\text{det}\ \mathcal{A}_4^{\phi^3}[1234|1234]&=-\frac{s_{13}}{s_{12}s_{14}}.
\end{align}
Thus there is one zero of the determinant $s_{13}=0$ and it is a physical pole. Due to the color-ordering constraints, consistency with locality requires that $\mathcal{A}_4[1234]$ does not have a pole at $s_{13}=0$. Thus, the zero of the determinant contributes a physical simple pole at $s_{13}=0$.

A similar structure exists at 5-point. Consider BCJ orderings like that in \cite{Bern:2019prr}, $[13524]$ and $[13542]$. This gives
\begin{align}
\text{det}\, \mathcal{A}_5^{\phi^3}[\alpha|\beta]&= -\frac{s_{23} s_{15}s_{34}}{s_{12} s_{13} s_{14} s_{24} s_{45}
	s_{35} s_{25}}.
\end{align}
Again we find that zeros of the determinant $s_{23} =s_{15}=s_{34}=0$, all correspond to physical poles. In addition, the color-ordering requires $\mathcal{A}_5[13524]$ and $\mathcal{A}_5[13542]$ to have no poles at these locations. Thus, zeros of the determinant contribute simple physical poles at 5-point order as well.

Requiring locality of the massless KLT formula at 4- and 5-point, provides another motivation for the fundamental BCJ relations. Only theories that satisfy the fundamental BCJ relations can be double-copied to local theories, whose amplitudes are free of non-physical poles and factorize on physical poles. 

Let us now investigate what happens to our proposed massive KLT formula at 4-point. We begin by choosing a basis of orderings $([1234],[1324])$. This gives,
\begin{align}
\det\, \mathcal{A}_4^{\phi^3}[\alpha|\beta]&=\frac{m^2}{(s_{12}+m^2)(s_{13}+m^2)(s_{14}+m^2)},
\end{align}
which has no zeros and thus no spurious pole can arise from the 4-point double-copy. This has the interesting consequence that any massive theory can be inserted into the massive KLT formula to obtain a 4-point amplitude of a local theory.

At 5-point, we are less lucky. Consider a basis of DDM orderings $[13\sigma(245)]$ where $\sigma$ runs over all 6 permutations of $(2,4,5)$, also used in \cite{Bern:2019prr}. Here we find
\begin{align}
\det \mathcal{A}_5^{\phi^3}[\alpha|\beta]=&\frac{m^8}{\prod_i\mathcal{D}_i}\mathcal{P}(s_{ij},m^2),
\end{align}
where
\begin{align}
\label{eq:spuriousdet}
\prod_i\mathcal{D}_i=&\left(m^2+s_{12}\right)^2 \left(m^2+s_{13}\right)^2 \left(m^2+s_{14}\right)^2 \left(m^2+s_{23}\right)^2 \left(m^2+s_{24}\right)^2\nonumber\\
& \left(m^2+s_{15}\right)^2 \left( m^2+s_{45}\right)^2 \left( m^2+s_{35}\right)^2\left(	m^2+s_{25}\right)^2 \left( m^2+s_{34}\right)^2,
\end{align}
and
\begin{align}
\mathcal{P}(s_{ij},m^2)=&\,320 m^8+36 m^6 (9 s_{12}+4 (s_{13}+s_{14}+s_{23}+s_{24}))\nonumber\\
&\hspace{-0.5cm}+m^4 \left(117 s_{12}^2+108 s_{12} (s_{13}+s_{14}+s_{23}+s_{24})+4 \left(s_{13} (13 s_{14}+4 s_{23}+17 s_{24})\right.\right.\nonumber\\
&\hspace{1cm}\left.\left.+4
s_{13}^2+4 s_{14}^2+17 s_{14} s_{23}+4 s_{14} s_{24}+4 s_{23}^2+13 s_{23} s_{24}+4
s_{24}^2\right)\right)\nonumber\\
&\hspace{-0.5cm}+2 m^2 \left(9 s_{12}^3+13 s_{12}^2 (s_{13}+s_{14}+s_{23}+s_{24})+s_{12} \left(s_{13} (10 s_{14}+6 s_{23}+17
s_{24})\right.\right.\nonumber\\
&\hspace{1cm}\left.\left.+4 s_{13}^2+4 s_{14}^2+s_{14} (17 s_{23}+6 s_{24})+2 (2 s_{23}+s_{24}) (s_{23}+2 s_{24})\right)\right.\nonumber\\
&\hspace{1cm}\left.+2 \left(s_{13}^2 (s_{14}+2 s_{24})+s_{13}
\left(s_{14}^2+s_{14} (s_{23}+s_{24})+s_{24} (s_{23}+2 s_{24})\right)\right.\right.\nonumber\\
&\hspace{1cm}\left.\left.+s_{23} \left(s_{24} (s_{14}+s_{23})+2 s_{14}
(s_{14}+s_{23})+s_{24}^2\right)\right)\right)\nonumber\\
&\hspace{-0.5cm}+2 s_{24} \left(s_{23} \left(s_{12}^2+s_{12} (s_{13}+s_{14})-s_{13} s_{14}\right)+s_{12}
(s_{12}+s_{13}) (s_{12}+s_{13}+s_{14})\right)\nonumber\\
&\hspace{-0.5cm}+(s_{12} (s_{12}+s_{13}+s_{14})+s_{23} (s_{12}+s_{14}))^2+s_{24}^2
(s_{12}+s_{13})^2.
\end{align}
Here, $\mathcal{D}_i$ contains all the physical poles and $\mathcal{P}$ is a quartic polynomial in Mandelstams. Allowing one of the Mandelstam invariants to vary independently, we find that there are four zeros of the determinant that \textit{do not} correspond to physical poles. As a result, unless the amplitudes $\mathcal{A}_5[13\sigma(245)]$ conspire to cancel these spurious poles, the proposed massive KLT formula will not give us amplitudes of a local theory. We expect that the presence of spurious poles will persist at higher-point.

This analysis of the equivalent KLT form of the proposed massive double-copy reveals a dangerous tension with locality. As we have argued, color-kinematics duality satisfying BCJ numerators exist (at least up to $n=5$) for generic models with uniform non-zero mass spectra. But such a double-copy will contain spurious singularities unless magical cancellations take place to remove them. Such cancellations will necessarily require additional relations among the DDM basis of partial amplitudes. Since there is no analogue of the usual BCJ relations, themselves a consequence of color-kinematics duality in massless models, these relations must be genuinely new constraints.

\section{Massive Gravity and \texorpdfstring{$\left(\text{Massive Yang-Mills}\right)^2$}{(Massive Yang-Mills)\texttwosuperior}}
\label{sec:mYM2}

To definitively establish that color-kinematics duality is not a sufficient condition for a double-copy to be physical, it is enough to construct a single explicit counterexample. In this section we analyze in detail the \textit{massive Yang-Mills} EFT and demonstrate that a BCJ representation of the scattering amplitudes with color-kinematics duality satisfying numerators exists, at least up to 5-point. We see that 3- and 4-point scattering amplitudes generated by the double-copy can be interpreted as coming from a theory of dRGT massive gravity and show that at 5-point the would-be double-copied amplitude contains spurious singularities. 

\subsection{Physical Motivation}
\label{sec:mYMhighE}
To understand the model we consider and the independent physical arguments that suggest a massive double-copy should be sensible, it is useful to begin with a slightly more general class of models. We consider models with a \textit{global} $U(N)$ symmetry with a spectrum of spin-1 states of mass $m$ transforming in the adjoint representation. To ensure the existence of a standard BCJ representation (\ref{mBCJform4}), we will restrict to interactions in which the color indices are contracted using only the (totally anti-symmetric) structure constants $f^{abc}$. The most general such model with parity-conserving interaction terms of mass dimension up to four is given by the Lagrangian\footnote{In this paper we will use the Lie algebra conventions $[T^a,T^b] = if^{abc}T^c$ and $\text{Tr}[T^a T^b]=\delta^{ab}$.}
\begin{equation}
\label{genericL}
    \mathcal{L} = -\frac{1}{4}\left(\partial_{[\mu}A^a_{\nu ]}\right)^2-\frac{1}{2}m^2 A_\mu^a A^{a\mu}-g f^{abc} A_\mu^a A_\nu^b \partial^\mu A^{c\nu} -\frac{1}{4} g'f^{abe}f^{cde}A_\mu^a A^{\mu c} A_\nu^b A^{\nu d}.
\end{equation}
Models of this kind with massive spinning states are generically only valid as low-energy effective descriptions. The associated scattering amplitudes violate perturbative unitarity bounds at a parametrically low energy scale unless special tunings of couplings are made or additional states such as Higgs bosons are introduced to \textit{soften} the UV behaviour. An efficient way to observe this is to study high-energy fixed angle, 2-to-2 scattering amplitudes. Here we use explicit center-of-mass frame kinematics with polarization vectors,  
\be
\begin{split}
\epsilon^{(\pm)}_{\mu}(p^i)&=(0,\mp\cos\theta^i,-i,\pm\sin\theta^i)\\
\epsilon^{(0)}_{\mu}(p^i)&=\frac{1}{m}(p,E\sin\theta^i,0,E\cos\theta^i),
\end{split}
\ee
and momenta
\be
p_\mu^i=(E,p\sin\theta^i,0,p\cos\theta^i),
\ee
with $i=1,2,3,4$ labeling the external particles scattering at angles $\theta^1=0,$ $\theta^2=\pi,$ $\theta^3=\theta,$ $\theta^4=\theta-\pi$. The worst behaved choice for the polarizations is given by purely longitudinal scattering\footnote{Here we are using a shorthand notation $\mathcal{A}\left(s_1 s_2 s_3 s_4\right)\equiv \mathcal{A}_4\left(1_{s_1}^{a_1},2_{s_2}^{a_2}\rightarrow 3_{s_3}^{a_3},4_{s_4}^{a_4} \right)$, where $a_i$ are adjoint indices and $s_i=+,-,0$ is the polarization.}
\begin{align}
\label{YMHE}
    \mathcal{A} \left(0000\right) &= \frac{1}{4m^4} \left(g^2-g'\right)\left[c_{12} \left(2 s^2+2 s t-t^2\right)+c_{13} \left(s^2-2 s t-2 t^2\right)\right] \nonumber\\
   &\hspace{4mm}+\frac{1}{4m^2}\left[c_{12} (4 g' (2 s+3 t)-g^2 (8 s+13 t))+c_{13} s (4 g'-3 g^2)\right]+\mathcal{O}\left(s^0\right),
\end{align}
where we have parametrized the expression in terms of the $m\rightarrow 0$ limit of the Mandelstam invariants 
\begin{equation}
    s \equiv 4E^2, \hspace{5mm} t \equiv 2E^2(\text{cos}(\theta)-1).
\end{equation}
We see that for generic values of $g'$ the scattering amplitudes grow like $E^4$ at high-energies, but for a specific tuning, $g'=g^2$, this is improved to $E^2$. If this tuning is made, the generic Lagrangian (\ref{genericL}) simplifies to
\begin{equation}
    \label{mYML}
    \mathcal{L} = -\frac{1}{4}\left(F_{\mu\nu}^a\right)^2 - \frac{1}{2}m^2 A_\mu^a A^{a\mu},
\end{equation}
where
\begin{equation}
    F^a_{\mu\nu} \equiv \partial_{[\mu}A^a_{\nu ]} + gf^{abc}A_\mu^b A_\nu^c,
\end{equation}
and defines the model we will study in this section under the name massive Yang-Mills. 

The improved high-energy behaviour of this tuning has a nice physical explanation. The massive Yang-Mills model has a simple (perturbative) UV completion as a particular limit of a Higgsed gauge theory. We begin with a model of scalar fields $\phi^{aa'}$ transforming in the bi-adjoint representation of $U(N)_L\times U(N)_R$ with a Higgs potential
\begin{equation}
    \mathcal{L}=-\frac{1}{2}\left(\partial_\mu \phi^{aa'}\right)^2 +\lambda v^2 \phi^{aa'}\phi^{aa'} - \frac{\lambda}{2}\left(\phi^{aa'}\phi^{aa'}\right)^2.
\end{equation}
When $\lambda >0$ and $v^2>0$, the $U(N)_L\times  U(N)_R$ symmetry is spontaneously broken to a $U(N)$ subgroup. Without loss of generality the vacuum expectation value can be taken to have the form
\begin{equation}
    \langle \phi^{aa'}\rangle = \frac{v}{N}\delta^{aa'},
\end{equation}
for which the unbroken subgroup $U(N)_V$ is generated by the ``vector-like" combinations\footnote{Here the adjoint generators are defined as $(T_L^i)^{ab}=f^{iab}$ and  $(T_R^i)^{a'b'}=f^{ia'b'}$.}
\begin{equation}
    (T_V^i)^{aa'bb'} = (T_L^i)^{ab} \delta^{a'b'}+ \delta^{ab}(T_R^i)^{a'b'}.
\end{equation}
If we gauge the orthogonal, broken ``axial-like" subgroup $U(N)_A$ generated by
\begin{equation}
   (T_A^i)^{aa'bb'} = (T_L^i)^{ab} \delta^{a'b'}- \delta^{ab}(T_R^i)^{a'b'},
\end{equation}
then in unitary gauge the associated $U(N)_A$ gauge bosons acquire masses $m_A\sim gv$, while preserving the unbroken global $U(N)_V$ symmetry under which they transform in the adjoint representation. The remaining $N^2(N^2-1)$ Higgs scalars have masses $m_H \sim \lambda^{1/2}v$, and in the limit $\lambda \rightarrow \infty$ with $v$ held fixed, decouple, with the low-energy dynamics of the massive vector bosons described by the massive Yang-Mills EFT.  

The Goldstone boson equivalence theorem \cite{PhysRevD.10.1145} tells us that the high-energy scattering of longitudinal vector modes of a spontaneously broken gauge theory must match the high-energy limit of a coset sigma model describing the same symmetry breaking pattern. In this case the coset is $(U(N)_L\times U(N)_R) / U(N)_V$, which is coincidentally the coset defining \textit{Chiral Perturbation Theory} ($\chi$PT) \cite{Weinberg:1978kz}, with the well-known Lagrangian 
\begin{equation}
    \mathcal{L} = \frac{f_\pi^2}{2}\text{Tr}\left[\partial_\mu U^\dagger \partial^\mu U\right], \hspace{5mm} U(x) \equiv \text{exp}\left(\frac{i}{f_\pi} T^a \pi^a(x)\right).
\end{equation}
The 2-to-2 scattering amplitude in this model is given by the simple expression
\begin{equation}
\label{ChPT4}
    \mathcal{A}_4\left(1,2,3,4\right)=\frac{1}{4 f_\pi^2} \left(-c_{12} t+c_{13} s\right),
\end{equation}
which precisely matches (\ref{YMHE}) in the limit $g'=g^2$, if the pion decay constant is identified as $f_\pi \sim m/g$. 

Massive Yang-Mills is not only a special EFT because it has softer than expected high-energy growth. As the above discussion indicates, in the high-energy limit the scattering amplitudes coincide with those of $\chi$PT, which is one of the few known massless models exhibiting color-kinematics duality \cite{Cachazo:2014xea,Chen:2013fya}, as can be verified explicitly using (\ref{ChPT4}). As a consequence, in the high-energy limit the massive Yang-Mills amplitudes can be double-copied to give the scattering amplitudes of the special Galileon \cite{Cheung:2014dqa},
\begin{equation}
    \left(\lim_{E \gg m} \mathcal{A}_n^{\text{mYM}}\right) \otimes \left(\lim_{E \gg m} \mathcal{A}_n^{\text{mYM}}\right) = \mathcal{A}_n^{\text{sGal}}.
\end{equation}
On the basis of this observation, it seems natural to conjecture the existence of some model of a massive spin-2 or \textit{massive gravity}, which matches the special Galileon amplitudes at high-energies and can be constructed as a double-copy 
\begin{equation}
    \mathcal{M}_n^{\text{mGrav}} \equiv \mathcal{A}_n^{\text{mYM}} \otimes_m \mathcal{A}_n^{\text{mYM}}. 
\end{equation}
An immediate problem with this is that we do not know what the symbol $\otimes_m$, denoting a massive double-copy, is supposed to mean. One property it should have, \textit{if} this story is self-consistent, is that it \textit{commutes} with the high-energy limit, meaning
\begin{equation}
    \label{commute}
    \lim_{E \gg m} \left(\mathcal{A}_n^{\text{mYM}}\otimes_m  \mathcal{A}_n^{\text{mYM}}\right) \stackrel{!}{=} \left(\lim_{E \gg m} \mathcal{A}_n^{\text{mYM}}\right) \otimes \left(\lim_{E \gg m} \mathcal{A}_n^{\text{mYM}}\right),
\end{equation}
where $\otimes$ on the right-hand-side is the familiar massless double-copy. In the Introduction (\ref{mBCJform4}), we described a natural generalization of the BCJ double-copy based on color-kinematics duality, to models with massive states, and in Section \ref{sec:massKLT} constructed an equivalent KLT-like formula. In this section we will demonstrate explicitly that such a double-copy does \textit{not} have the property (\ref{commute}) and moreover, for $n>4$ does not produce a physical scattering amplitude that can be matched to a local Lagrangian.

\subsection{3-point Amplitudes and Asymptotic States}
\label{sec:3amps}

Before considering the dynamical content of the double-copy, we first need to understand the mapping of states in the asymptotic Hilbert space. Massive Yang-Mills is a model of a massive vector boson, with 3 on-shell degrees of freedom in $d=4$. The Hilbert space of asymptotic one-particle states is spanned by the space of plane-wave solutions to the linearized equations of motion. In the present context it is convenient to represent the basis of linearly independent plane-wave solutions using the massive spinor formalism of \cite{Arkani-Hamed:2017jhn}. In this approach, the 3 independent spin states are collected together into a rank-2, totally symmetric $SU(2)$ little group tensor. Explicitly,
\begin{equation}
\label{vpw}
    A_\mu^{a\;IJ}(x) = c^a\epsilon^{IJ}_\mu(p)e^{ip\cdot x}, \hspace{5mm} \text{where} \hspace{5mm}\epsilon^{IJ}_\mu(p) = -\frac{1}{2\sqrt{2}}\tilde{\lambda}_{\dot{\alpha}}^{(I}\overline{\sigma}^{\dot{\alpha}\alpha}_\mu\lambda^{J)}_\alpha.
\end{equation}
The double-copy of such a plane-wave solution is given simply by replacing the color factor $c^a$ with a second copy of the polarization vector,
\begin{equation}
\label{fatgrav}
    A^{a\;IJ}_\mu(x) \otimes A^{b\; KL}_\nu(x) = \mathfrak{h}_{\mu\nu}^{IJKL}(x) \equiv \epsilon^{IJ}_\mu(p)\epsilon^{KL}_\nu(p)e^{ip\cdot x}.
\end{equation}
Where (\ref{vpw}) transforms in an irreducible representation of $SU(2)$, the double-copy (\ref{fatgrav}) transforms in a reducible representation. Such a plane-wave double-copy is equivalent to a tensor product of one-particle Hilbert spaces, for which standard decomposition of representations of $SU(2)$ gives the physical spectrum of the double-copy
\begin{equation}
    \mathbf{3}\otimes \mathbf{3} = \mathbf{5}\oplus \mathbf{3} \oplus \mathbf{1}.
\end{equation}
Hence we expect the double-copy of massive Yang-Mills to describe a model of a massive \textit{graviton} $h_{\mu\nu}$ (spin-2) coupled to a massive \textit{Kalb-Ramond two-form} $B_{\mu\nu}$ (spin-1) and a massive \textit{dilaton} $\phi$ (spin-0). It is most convenient to first calculate the scattering amplitudes for the reducible $\mathfrak{h}$-states, and project out the physical states as needed. To extract the physical spectrum of the double-copy we use the following \textit{projection operators}\footnote{The normalization constants can be fixed by requiring that the completeness relation for polarizations gives the same sum over states before and after projecting onto physical states.}
\begin{align}
  (P_h)_{I_1 I_2 J_1 J_2}^{K_1 K_2 K_3 K_4} &= \frac{1}{24} \delta^{(K_1 K_2 K_3 K_4)}_{I_1 I_2 J_1 J_2}, \hspace{1.5mm} (P_B)_{I_1 I_2 J_1 J_2}^{K_1 K_2} = \frac{1}{\sqrt{2}} \epsilon_{I_1 J_1}\delta_{I_2 J_2}^{(K_1 K_2)}, \nonumber\\
  &\hspace{5mm}(P_\phi)_{I_1 I_2 J_1 J_2} = \frac{1}{\sqrt{3}} \epsilon_{I_1 J_1}\epsilon_{I_2 J_2}.
\end{align} 
The physical polarization tensor of the two-form is antisymmetric $\epsilon^{(B)}_{\mu\nu} = -\epsilon^{(B)}_{\nu\mu}$, and consequently gives a non-vanishing contribution to amplitudes in the double-copy only if there are an even number of such states. Equivalently, the two-form has a $\mathds{Z}_2$ symmetry, which allows us to form a consistent truncation containing only the graviton and dilaton modes.

Since the polarization tensors in the truncated model are symmetric we can represent the amplitudes using a convenient shorthand. We suppress the little-group indices by making the replacement $\epsilon^{I_i I_i}_\mu(p_i) \rightarrow z_\mu^i$; the amplitude is then a rational function of the following elementary building blocks:
\begin{equation}
    p_{ij} \equiv p^i_{\mu}{p^j}^\mu, \hspace{5mm} z_{ij} \equiv z^i_{\mu}{z^j}^\mu, \hspace{5mm} zp_{ij} \equiv z^i_{\mu}{p^j}^\mu.
\end{equation}
  Extracting the physical graviton and dilaton states amounts to the replacement rules,
\begin{align}
    \label{reprules}
    z^i_\mu z^i_\nu &\rightarrow \epsilon_{\mu\nu}(p_i) &&\text{(Massive Graviton)} \nonumber\\
    z^i_\mu z^i_\nu &\rightarrow \frac{1}{\sqrt{3}}\bigg(\eta_{\mu\nu}+\frac{p_{i\mu} p_{i\nu}}{m^2}\bigg) &&\text{(Massive Dilaton)}.
\end{align}
We begin with the double-copy of 3-point scattering amplitudes. This is of course unconstrained by color-kinematics duality, but will be important for reconstructing the massive gravity Lagrangian from the 4-point amplitudes. A local BCJ representation of the massive Yang-Mills amplitudes can be efficiently constructed using the Feynman rules given in Appendix \ref{app:Feyn}. The cubic Yang-Mills amplitude is given by 
\be
\label{eq:cubicnot}
\mathcal{A}_3=2g\big(z_{23}zp_{12}+z_{13}zp_{23}+z_{12}zp_{31}\big).
\ee
The gravitational amplitude is given by squaring the Yang-Mills amplitude and replacing the coupling constants as $g^2\rightarrow \frac{1}{2 M_p}$, giving
\be
\label{eq:cubicgrav}
\mathcal{M}_3=\frac{2}{M_p}\big(z_{23}zp_{12}+z_{13}zp_{23}+z_{12}zp_{31}\big)^2.
\ee
Using (\ref{reprules}) we can extract from this the cubic amplitudes for physical states. The on-shell cubic amplitude for 3 gravitons is formally identical to the massless case, given by:
\begin{align}
 \mathcal{M}(1_h,2_h,3_h)=   \frac{2}{M_p}\bigg( &{\epsilon_1}_{\mu\nu} {\epsilon_2}^{\mu\nu} {\epsilon_3}_{\alpha\beta}{p_1}^\alpha {p_1}^\beta+2\ {p_2}^\mu{\epsilon_1}_{\mu\nu} {\epsilon_2}^{\nu\alpha}{\epsilon_3}_{\alpha\beta} {p_1}^{\beta}\nonumber\\
 &+\text{cyclic permutations of}\ (1,2,3)\bigg).
\end{align}

The amplitude for 2 gravitons and 1 dilaton is given by
\be
\label{eq:grav2dil1}
\mathcal{M}_3\left( 1_h, 2_h, 3_\phi \right)=-\frac{\sqrt{3}}{2M_p} m^2 {\epsilon_1}_{\mu\nu} {\epsilon_2}^{\mu\nu}.
\ee
We see that this expression vanishes as $m\rightarrow 0$, recovering the expected massless amplitude. It is interesting to note that the $\mathds{Z}_2$ \textit{dilaton parity} of the massless double-copy only emerges in the massless limit. Therefore when $m\neq0$ we cannot make a further consistent truncation to the gravity sector. The on-shell cubic amplitudes for 1 graviton and 2 dilatons is given by
\bea
\label{eq:grav1dil2}
\mathcal{M}_3\left( 1_h, 2_\phi, 3_\phi \right)=\frac{3}{2M_p} {\epsilon_1}_{\mu\nu}{p_2}^\mu {p_2}^\nu.
\eea
This vertex appears in both the massive and massless cases. The on-shell cubic amplitude for 3 dilatons is given by
\be
\label{eq:cubicdil}
\mathcal{M}_3\left( 1_\phi, 2_\phi, 3_\phi \right)=-\frac{11\sqrt{3}}{8M_p} m^2.
\ee
This cubic dilaton vertex is also unique to the massive case and does not appear in the massless case. 

\subsection{4-point Amplitudes and High Energy Behavior\label{quartic}}
A BCJ representation of the 4-point amplitude is straightforwardly generated from the Feynman rules in Appendix \ref{app:Feyn}. This gives the following massive kinematic numerators
\be
\begin{split}
\label{num4s}
n_{12}=&[(\epsilon_1\cdot \epsilon_2)p_1^\mu+2(\epsilon_1\cdot p_2)\epsilon_2^{\mu}-(1\leftrightarrow 2)]\bigg(g_{\mu\nu}+\frac{({-p_1}_{\mu}-{p_2}_{\mu})({p_3}_{\nu}+{p_4}_{\nu})}{m^2}\bigg)\\
&\times[(\epsilon_3\cdot \epsilon_4){p_3}^\nu+2(\epsilon_3\cdot p_4){\epsilon_4}^{\nu}-(3\leftrightarrow 4)]\\
&+(s+m^2)[(\epsilon_1\cdot\epsilon_3)(\epsilon_2 \cdot \epsilon_4)-(\epsilon_1\cdot\epsilon_4)(\epsilon_2 \cdot \epsilon_3)],
\end{split}
\ee
with the first two lines coming from the exchange diagrams and the third line coming from the contact diagram. The other numerators are found by taking
\be
\label{num4tu}
n_{13} = n_{12} |_{1\rightarrow 3\rightarrow 2\rightarrow 1},\quad n_{14} = n_{12} |_{1\rightarrow 2\rightarrow3\rightarrow1}.
\ee
The $1/m^2$ term in the massive vector propagator vanishes, and so these numerators are formally identical to the Feynman rule-generated expressions for massless Yang-Mills. As a surprising consequence of the fact that at 4-point, all generalized gauges satisfy the kinematic Jacobi identity, we find that likewise for (\ref{num4s}) and (\ref{num4tu})
\begin{equation}
    n_{12}+n_{13}+n_{14}=0.
\end{equation}
The 4-point massive gravity amplitude is then given by
\be
\mathcal{M}_4=\frac{1}{4 M_p^2}\bigg(\frac{n^2_{12}}{s_{12}+m^2}+\frac{n^2_{13}}{s_{13}+m^2}+\frac{n^2_{14}}{s_{14}+m^2}\bigg).
\ee
The explicit expressions for the physical scattering amplitudes are rather complicated and are given explicitly in Appendix \ref{app:4amps}\footnote{These results are in agreement with those that appeared recently in \cite{Momeni:2020vvr}.}.

We expect the double-copy procedure for massive Yang-Mills to give a ghost-free theory of massive gravity\footnote{See \cite{Hinterbichler_2012} for a review of massive gravity.}. Generic ghost free massive gravity without coupling to a dilaton, also known as dRGT, propagates 5 degrees of freedom, has two free parameters in $D=4$, and is given by the action
\begin{equation}
S=\frac{M_{P}^{D-2}}{2}\int d^D x\Big[(\sqrt{-g}R)-\sqrt{-g}\frac{1}{4}m^2W(g,\mathcal{K})\Big],
\end{equation}
where 
\begin{equation}
W(g,\mathcal{K})=\sum_{n=2}^{n=D}\alpha_n \mathcal{L}^{TD}_n (\mathcal{K}),
\end{equation}
brackets mean trace with respect to the full metric, $\alpha_2=-4$, and the rest of the coefficients are arbitrary \cite{de_Rham_2011, Hassan:2011hr}. The tensor $\mathcal{K}^\mu_\nu(g,H)$ is given by
\begin{equation}
\mathcal{K}^{\mu}_\nu=\delta^{\mu}_\nu-\sqrt{\delta^{\mu}_\nu-H^{\mu}_\nu}=\sum_{n=1}^\infty d_n(H^n)^\mu_\nu, \quad d_n=-\frac{(2n)!}{(1-2n)(n!)^2 4^n},
\end{equation}
where indices are raised by the full metric $g^{\mu}_\nu=\gamma^{\mu}_\nu+h^{\mu}_\nu$, the background metric is $\gamma^{\mu}_\nu$, and $H^{\mu}_\nu=g^{\mu}_\nu-\tilde{\gamma}^{\mu}_\nu$ is the St{\"u}ckelberg replacement for $h^{\mu}_\nu$. The quantity $\mathcal{L}^{TD}_n(\Pi)$ can be written as total derivatives when $\Pi=\partial_{\mu} \partial_{\nu} \phi$. These total derivative combinations are unique up to an overall constant and can be found using the recursion relation
\begin{equation}
\mathcal{L}^{TD}_n(\Pi)=-\sum_{m=1}^n(-1)^m \frac{n!}{(n-m)!}\Pi^m_{\mu \nu} \mathcal{L}^{TD}_{n-m}(\Pi),
\end{equation}
with $\mathcal{L}^{TD}_0=1$. 

Massive gravity with the most generic potential without the dRGT tuning has an extra scalar degree of freedom that is ghostly and 4-point scattering amplitudes that grow with center-of-mass energy like $E^{10}$. However, the dRGT tuning, which leaves only 2 free parameters, removes the ghostly degree of freedom and improves the high energy behavior to scale with energy as $E^{6}$ ~\cite{Cheung:2016yqr,Bonifacio:2017iry}. Another common parameterization of dRGT massive gravity is given in \cite{de_Rham_2010}. The leading high energy behavior in this parameterization, for the tree-level 4-point amplitude for dRGT massive gravity, is given by:
\bea
\label{eq:highE}
\mathcal{M}(1^+1^+1^+1^+) &=& -\frac{3}{32} (1 - 4c_3) s^3\\
 \mathcal{M}(1^+1^+1^-1^-) &=& \mathcal{M}(1^-1^-1^+1^+) = \frac{9}{32} (1 - 4c_3)^2 st(s + t) \\
  \mathcal{M}(2^+000)&=&\frac{1}{\sqrt{6}}(c_3 +8d_5)st(s+t)\\
\label{eq:highE3355}  
\mathcal{M}(1^+1^+00)&=& \frac{1}{32}s\Big(2\big(1-8c_3+48c_3^2+64d_5)t(s+t)-3(1-4c_3)^2s^2\Big)\\
\mathcal{M}(1^+1^-00)&=&\frac{1}{96}s\big((1+12c_3)^2+384d_5\big)st(s+t)  \\
\label{eq:highEend}
\mathcal{M}(0000) &=& \frac{1}{6}\big(1 + 4c_3(9c_3-1)+64d_5\big) st(s + t),
\eea
where the polarization tensors, $\epsilon^{(a)}_{\mu\nu}$, have been split into two tensor modes $(a=2^+,2^-)$, two vector modes $(a=1^+,1^-)$, and one scalar mode $(a=0)$, the relation between the free parameters of dRGT are given by:
\be
\alpha_3=-2c_3\ \text{ and }\ \alpha_4=-4 d_5,
\ee
and the polarization tensors are chosen to be:
\be
\label{eq:gravpol}
\begin{split}
\epsilon^{(2\pm)}_{\mu\nu}&=\epsilon^{(\pm)}_\mu\epsilon^{(\pm)}_\nu,\\
\epsilon^{(1\pm)}_{\mu\nu}&=\frac{1}{\sqrt{2}}\bigg(\epsilon^{(\pm)}_\mu\epsilon^{(0)}_\nu+\epsilon^{(0)}_\mu\epsilon^{(\pm)}_\nu\bigg)\\
\epsilon^{(0)}_{\mu\nu}&=\frac{1}{\sqrt{6}}\bigg(\epsilon^{(+)}_\mu\epsilon^{(-)}_\nu+\epsilon^{(-)}_\mu\epsilon^{(+)}_\nu+2\epsilon^{(0)}_\mu\epsilon^{(0)}_\nu\bigg).
\end{split}
\ee
Indeed the 3-point amplitude (\ref{eq:cubicgrav}) corresponds to dRGT massive gravity with
\be
\alpha_3=-\frac{1}{2}\ \text{ or }\ c_3=\frac{1}{4}. 
\ee
This value is also the one picked out in the eikonal approximation analysis needed to avoid superluminal propagation as shown in \cite{Bonifacio:2018vzv} and is the ``partially massless" $\alpha_3$ \cite{deRham:2018svs,de_Rham_2011}. 

With the new cubic vertices that appear in the massive case, there are new scattering channels that appear in the quartic amplitudes that would not appear in the massless case. In agreement with the general discussion in Section \ref{sec:factor}, we find that all quartic amplitudes factorize properly on the poles into products of the corresponding 3-point amplitudes. For example, in the 4-graviton scattering amplitude, we find contributions from diagrams corresponding to the $s,t,u$ channels mediated by both a massive graviton and a dilaton, due to the non-vanishing cubic coupling with 2 gravitons and 1 dilaton. The 4-graviton amplitude matches that of massive gravity with the coefficients 
\be
\alpha_4=\frac{7}{48}\ \text{ or }\ d_5=-\frac{7}{192},
\ee
plus the additional channels mediated by the dilaton. 

At first glance, it may appear that a field redefinition could mix the cubic $hh\phi$ vertex and massive gravity quartic interactions, leading to the choice of $\alpha_4$ to not be uniquely specified. Since amplitudes are unaffected by field redefinition, we consider the difference between the double-copied amplitude and the dRGT massive gravity amplitude with $\alpha_3=-\frac{1}{2}$ and $\alpha_4$ left unspecified. We find terms proportional to $\sim(48\alpha_4-7)Tr[\epsilon_1\cdot\epsilon_2\cdot\epsilon_3\cdot\epsilon_4]$. This structure cannot be altered by introducing scalar channel diagrams and thus, requiring that it vanish picks out the remaining parameter to be $\alpha_4=\frac{7}{48}$.

The leading high energy behavior of the amplitudes for graviton-graviton scattering in the massive double-copy goes as:
\bea
\label{mgHE}
  \mathcal{M}(2^+000)&=&-\frac{1}{24\sqrt{6}}st(s+t)\\
\mathcal{M}(1^+1^+00)&=&- \frac{1}{48}st(s+t)\\
\mathcal{M}(1^+1^-00)&=&\frac{1}{48}st(s+t)  \\
\mathcal{M}(0000) &=& \frac{7}{144}st(s + t).
\eea
For the value of $c_3$ picked out by the double-copy, the high energy behavior of the 4-point amplitudes for massive gravity, \eqref{eq:highE} through \eqref{eq:highEend}, is improved for amplitudes where all the polarizations of the external particles are vector modes, scaling as $E^4$ rather than $E^6$. The dilaton affects the coefficient of $\mathcal{M}(0000)$, the amplitude where all the external particles are scalar modes. Without the dilaton, this amplitude would behave as $ \mathcal{M}(0000) = -\frac{1}{72}st(s + t)$. All other amplitudes behave as they would without the dilaton and are consistent with the above $c_3$ and $d_5$ values in expressions \eqref{eq:highE} through \eqref{eq:highEend}. 

One immediate and important result from (\ref{mgHE}) is that the conjectured property (\ref{commute}) does not hold for the BCJ double-copy. In the Goldstone boson equivalence limit for massive Yang-Mills, only the spin-1 longitudinal mode contributes at $E^2$. If (\ref{commute}) held, we would expect only the scattering of a single scalar mode to contribute at $E^6$ in the double-copy. From (\ref{mgHE}), we see explicitly that this is not the case.

In the 4-point amplitude where all the external particles are dilatons, there will be $s,t,u$ channels mediated by a massive graviton, as well as $s,t,u$ channels mediated by a dilaton, and a 4-dilaton contact term. The massless case only has the channels mediated by the massless graviton. 

The 4-point amplitude with 2 gravitons and 2 dilatons exists in the massless and massive case.  In the massless case, this 4-point amplitude has graviton exchange channels via the $h\phi\phi$ and $hhh$ vertices and dilaton exchange channels via two $h\phi\phi$ vertices, plus a contact term $hh\phi\phi$. In the massive case, there will be additional graviton exchange channels via two $hh\phi$ vertices, as well as dilaton exchange channels via the vertices $hh\phi$ and $\phi\phi\phi$.

The 4-point amplitudes with 3 gravitons and 1 dilaton or 1 graviton and 3 dilatons are unique to the massive case and involve all possible exchange diagrams with dilaton propagators, as well as graviton propagators, and with additional $hhh\phi$ and $h\phi\phi\phi$ contact terms. 

 The high energy behavior of all the amplitudes scales with energy like $\sim E^6$ or less and the amplitudes that scale like $E^6$ take the special galileon form $s t(s+t)$ \cite{Cheung:2014dqa}. As another example, the leading high energy behavior of $h\phi\phi\phi$ amplitudes is shown below:
\bea
\mathcal{M}(2^+\phi\phi\phi) &=& -\frac{s t (s+t)}{96\sqrt{3}}\\
\mathcal{M}(0\phi\phi\phi) &=& -\frac{11s t (s+t)}{288 \sqrt{2}}.
\eea
All the 4-point graviton and dilaton amplitudes resulting from the double-copy are given in Appendix \ref{app:4amps}.

	\subsection{5-point Amplitudes and Non-Physical Singularities\label{quintic}}
	As discussed in Section \ref{sec:factor}, 5-point massive gravity amplitudes constructed via the massive KLT formula are guaranteed to factorize correctly into 4- and 3-point amplitudes (listed in Appendix \ref{app:4amps} and Section \ref{sec:3amps} respectively). Nonetheless as we saw at 4-point, checking factorization at 5-point is a good cross-check of our more general results, in particular those of Appendix \ref{app:factor}.
	
	We begin by choosing a DDM basis of orderings $[13\;\sigma(2,4,5)]$ where $\sigma$ runs over all possible permutations. Using the Feynman rules of massive Yang-Mills, we then calculate partial amplitudes and use the inverse of bi-adjoint scalar matrix \eqref{eq:biaj5pt} to construct 5-point all-graviton amplitudes. The inverse of \eqref{eq:biaj5pt} is unwieldy so we do the following numerical tests of factorization.
	
	One can choose an independent basis of building blocks from the set of all $(\epsilon_i\cdot\epsilon_j)$, $(\epsilon_i\cdot p_j)$ and $(p_i\cdot p_j)$. We then assign numeric values to all these kinematic structures except one, without loss of generality let's call this $(p_1\cdot p_2)$. One can then evaluate the 5-point amplitude on this set of kinematic data and check that the residue on physical pole $(p_1\cdot p_2)=-\frac{m^2}{2}$ is exactly what one would expect
	\begin{equation}
	 \underset{s_{12}=-m^2}{\text{Res}} \mathcal{M}_5(12345)\overset{!}{=}\sum_X\mathcal{M}_3\left(1 2 (-P_{12})_X\right)\times\mathcal{M}_4\left((P_{12})_{\bar{X}}3 4 5\right),
	\end{equation}
	where $X$ can either be a dilaton or a graviton. As expected from the general discussion in Section \ref{sec:factor} we find that the would-be 5-point amplitude factors as expected on physical poles.
	
	While the correct factorization of the 5-point amplitude is promising, we saw in Section \ref{sec:spurious} that the KLT kernel suffers from non-physical poles arising from the determinant of the matrix of bi-adjoint scalar amplitudes. These singularities \eqref{eq:spuriousdet} can only be removed if special cancellations occur between amplitudes in the theory we are double-copying and the KLT kernel.
	
	In the context of this explicit example, we can proceed with our numerical analysis to check for example, whether all poles in $(p_1\cdot p_2)$ are physical. This can be done by evaluating the KLT formula on an incomplete set of kinematic data that leaves $(p_1\cdot p_2)$ unspecified. One can then check if all singularities in $(p_1\cdot p_2)$ are accounted for by locality. We find that this is not the case and that the resulting 5-point amplitude  $\mathcal{M}_5$ \textit{does have spurious poles}. The singularity structure takes exactly the form \eqref{eq:spuriousdet} which can be recast as
	\begin{align}
		\mathcal{P}(s_{ij},m^2)=&\alpha_1 s_{12}^4+\alpha_2 s_{12}^3+\alpha_3 s_{12}^2+\alpha_4 s_{12}+\alpha_5\,,
	\end{align}
	where $\alpha_i$ are functions of the mass and other Mandelstam variables. Since this polynomial does not easily factor into rational roots, it is useful to choose special kinematic configurations where it factors more readily. In these cases, the exact locations of the spurious poles can be found and the amplitude evaluated on such a non-physical pole gives a nonzero residue.
	
	Thus, no miraculous cancellations occur in massive Yang-Mills to get rid of spurious singularities. In particular this means that in its current form, \textit{massive Yang-Mills does not sensibly double-copy to massive gravity}.
	
	Furthermore, if we attempt to save the double-copy, by for example, adding a 5-point contact contribution to cancel these non-physical poles, we find no improvement. Consider for example adding a new operator at 5-point, such that
	\begin{align}
	\tilde{\mathcal{A}}_5[13542]&=\mathcal{A}_5[13245]+\frac{\alpha g^3}{m^2} (p_1\cdot\epsilon_3)(\epsilon_1\cdot\epsilon_2)(\epsilon_3\cdot\epsilon_5),
	\end{align}
	with contributions to the other orderings determined by relabeling. Here $\alpha$ is a free coefficient. The powers of $m^2$ have been introduced to correct the mass dimension, this would correspond to adding a term $\sim\partial A^5$ to the massive Yang-Mills Lagrangian.
	
    We find that there is \textit{no way} to tune $\alpha$ to remove any of the spurious singularities. Since it is unclear whether this statement still holds for arbitrary combinations of the other 28 possible $\partial A^5$ structures, we cannot strictly rule out the possibility of a massive Yang-Mills 5-point operator removing non-physical poles from the KLT product. Nonetheless, our calculation is indicative that this may not be possible.

\section{Locality and the Spectral Condition}
\label{sec:diffmass}

We have seen that the proposed massive KLT construction \eqref{eq:KLTmassive} is in serious tension with locality. In general, the inverse of the matrix of KK independent massive bi-adjoint scalar amplitudes contains spurious, non-physical singularities \eqref{eq:spuriousdet}. For the full KLT sum to be free of these non-physical singularities, additional non-trivial constraints must be imposed. These conditions are not met in the case of massive Yang-Mills, because as we saw in Section \ref{quintic}, the resulting 5-point massive gravity amplitude is not local. Thus, despite the existence of color-kinematics duality satisfying numerators for all KK satisfying models, the resulting would-be double copies only correspond to physical amplitudes if additional constraints are imposed. 
	
To better understand these additional constraints, let us first look at the massless case. Here the additional constraints are the fundamental BCJ relations and color-kinematics duality satisfying numerators can only be found in theories whose amplitudes are BCJ-compatible. In the language of bi-adjoint scalar theory, the double-copy formulation gives rise to physical amplitudes only if the $(n-2)!\times (n-2)!$ matrix of bi-adjoint scalar amplitudes $\mathcal{A}^{\phi^3}[\alpha|\beta]$ has rank $(n-3)!$, which we will refer to as \textit{minimal rank}. In addition, only theories whose amplitudes satisfy the fundamental BCJ relations, which arise as null vectors of the singular matrix of bi-adjoint scalar amplitudes, can be double-copied.
	
In the massive case, a matrix of bi-adjoint scalar amplitudes that has minimal rank can be constructed if a specific condition on the masses, given by the equation $\det \mathcal{A}^{\phi^3}[\alpha|\beta]=0$ is met. We will call this the \textit{spectral condition}. The null vectors of this matrix will then give rise to massive BCJ relations. On the basis of this observation, we propose the following:

\textbf{Conjecture:} \textit{The KLT prescription for double-copying models with massive states generates physical amplitudes without spurious singularities, and reduces smoothly to the massless double-copy in an appropriate $m\rightarrow 0$ decoupling limit, if the associated bi-adjoint scalar matrix has minimal rank.}

In this section we will illustrate the consequences of imposing these conditions on models at $n=4$ and $n=5$. We will see how this alternative construction has both a commuting decoupling limit and the absence of spurious singularities, providing evidence in support of our conjecture above.  
	
\subsection{4-point Spectral Condition}
We will begin with a model that has a more general spectrum of massive or massless states. We denote the external states $m_i$ and the intermediate masses being exchanged on a factorization channel as $m_{ij}$. The only assumption we will make is the existence of a BCJ representation of the form
	\begin{equation}
	\mathcal{A}_4\left(1^{a_1},2^{a_2},3^{a_3},4^{a_4}\right) = \frac{c_{12}n_{12}}{s_{12}+m_{12}^2}+ \frac{c_{13}n_{13}}{s_{13}+m_{13}^2} +\frac{c_{14}n_{14}}{s_{14}+m_{14}^2}.
	\end{equation}
Implicitly built into this expression is the assumption that only states with mass $m_{12}^2$ are exchanged in the $s_{12}$-channel and so forth. This is not completely general and an interesting open problem is to construct an appropriate generalization of the BCJ form for models with multiple mass states exchanged in a single channel. We now choose a DDM basis $([1234], [1324])$, in which the matrix of bi-adjoint scalar amplitudes is
	\begin{equation}
	\mathcal{A}_4^{\phi^3}[\alpha|\beta] = 
	\begin{pmatrix}
	\frac{1}{s_{12}+m_{12}^2} +\frac{1}{s_{14}+m_{14}^2} & - \frac{1}{s_{14}+m_{14}^2} \\
	- \frac{1}{s_{14}+m_{14}^2} & \frac{1}{s_{13}+m_{13}^2} +\frac{1}{s_{14}+m_{14}^2}
	\end{pmatrix}.
	\end{equation}
Taking the determinant, gives
	\begin{equation}
	\text{det}\;\mathcal{A}_4^{\phi^3} = \frac{m_{12}^2+m_{13}^2+m_{14}^2-m_1^2-m_2^2-m_3^2-m_4^2}{(s_{12}+m_{12}^2)(s_{13}+m_{13}^2)(s_{14}+m_{14}^2)}.
	\end{equation}
Clearly $\mathcal{A}^{\phi^3}[\alpha|\beta]$ is full-rank and non-singular, i.e. $\det \mathcal{A}^{\phi^3}[\alpha|\beta]$ does not vanish, for generic mass spectra. In keeping with our conjecture, we want to reduce the rank of $\mathcal{A}^{\phi^3}[\alpha|\beta]$ to (4-3)!=1, which is the minimal rank at 4-point order. This is achieved by imposing the following condition on the mass spectrum of the theory,
	\begin{equation}
	\label{eq:speccond}
	m_{12}^2+m_{13}^2+m_{14}^2 = m_1^2+m_2^2+m_3^2+m_4^2.
	\end{equation}
This is the \textit{4-point spectral condition}. It is interesting to note that the spectrum of massive Yang-Mills does not satisfy this condition. We will see later that this is what led the double-copy and decoupling limit to fail to commute when studying massive $(\text{Yang-Mills})^2$.
	
On imposing the spectral condition, $\mathcal{A}^{\phi^3}[\alpha|\beta]$ becomes singular and is no longer invertible. As a result, we must eliminate one row and one column to produce an invertible matrix of bi-adjoint scalar amplitudes. This is consistent only if all such choices give the same result. For example, we could remove the second row and second column, the resulting KLT formula is then
\begin{equation}
\label{eq:KLT4}
\mathcal{M}_4\left(1,2,3,4\right) = -\frac{(s_{12}+m_{12}^2)(s_{14}+m_{14}^2)}{s_{13}+(m_1^2+m_2^2+m_3^2+m_4^2-m_{12}^2-m_{14}^2)}\,\mathcal{A}_4[1,2,3,4]^2.
\end{equation}
	If however, we choose to eliminate the second row and the first column, we find 
\begin{equation}
\mathcal{M}_4\left(1,2,3,4\right) = -(s_{14}+m_{14}^2)\mathcal{A}_4[1,2,3,4]\mathcal{A}_4[1,3,2,4].
\end{equation}
Equating these formulae constructs a \textit{massive} version of the fundamental BCJ relation
\begin{equation}
\label{eq:massiveBCJ4}
(s_{12}+m_{12}^2)\mathcal{A}_4[1,2,3,4] = (s_{13}+m_{13}^2)\mathcal{A}_4[1,3,2,4],
\end{equation}
where we have used the spectral condition to rewrite the relation in a more compact form.

As we prove in Appendix \ref{app:BCJnull}, an equivalent way to derive the massive BCJ relation is by studying the null vector of $\mathcal{A}^{\phi^3}[\alpha|\beta]$ which is
\begin{align}
	\vec{n}&=\begin{pmatrix}
	-s_{12}-m_{12}^2\\
	s_{13}+m_{13}^2
	\end{pmatrix}\,.
\end{align}
Setting the dot product of this vector with the DDM basis to zero then gives the BCJ relation,
\begin{align}
	\vec{n}\cdot \left(\mathcal{A}_5[1234]\hspace{0.5cm} \mathcal{A}_5[1324]\right)&=(s_{12}+m_{12}^2)\mathcal{A}_4[1,2,3,4] - (s_{13}+m_{13}^2)\mathcal{A}_4[1,3,2,4]=0.
\end{align}

We are now in a position to study the singularity structure of the KLT formula \eqref{eq:KLT4}. The first aspect of the formula that we note is the absence of spurious poles, i.e. all poles are at physical locations. To ensure locality, we can study the amplitude in the neighbourhood of its three physical poles. For example,
\begin{align}
	\underset{s_{12}=-m_{12}^2}{\text{Res}}\mathcal{M}_4\left(1,2,3,4\right)&=-\frac{(s_{14}+m_{14}^2)}{s_{13}+m_{13}^2}\mathcal{A}_4[1,2,-P_{12}]^2\mathcal{A}_3[P_{12},3,4]^2\nonumber\\
	&=\mathcal{A}_4[1,2,-P_{12}]^2\mathcal{A}_3[P_{12},3,4]^2\nonumber\\
	&=\mathcal{M}_3\left(1,2,-P_{12}\right)\mathcal{M}_3\left(P_{12},3,4\right)\,,
\end{align}
where we have used $s_{13}+m_{13}^2=-s_{14}-m_{14}^2$ on the $s_{12}$ pole. Thus the amplitude factorizes correctly on the $s_{12}$ pole. Factorization on the $s_{13}$ and $s_{14}$ pole follow in a similar manner.

It is easy to see that these forms of the massive BCJ relations and KLT formula smoothly reduce to the massless ones when all external and intermediate masses, $m_i$ and $m_{ij}$ are taken to zero. As a result, this version of the massive double-copy \textit{does commute with the decoupling limit}. Thus for any pair of massive BCJ-compatible theories $A^{(m)}$ and $B^{(m)}$ that satisfy the spectral condition, one can construct a local theory,
\begin{align}
	C^{(m)}&=A^{(m)}\otimes_m B^{(m)}\,,
\end{align}
where $\otimes_m$ is our conjectured massive KLT formalism. This will reduce in the decoupling limit to
\begin{align}
	\lim_{m\to0}C^{(m)}&=\lim_{m\to0}\left(A^{(m)}\otimes_m B^{(m)}\right)=\left(\lim_{m\to0}A^{(m)}\right)\otimes \left(\lim_{m\to0}B^{(m)}\right)\,,
\end{align}
where $\otimes$ denotes the massless KLT double-copy.

As we saw in Section \ref{sec:mKLT}, the massive KLT and massive BCJ double copies are equivalent. Let us now understand our conjecture from the perspective of the BCJ double-copy. We begin by considering the effect of a generalized gauge transformation on the BCJ representation, similar to \eqref{eq:ggt}. The amplitude is invariant under the following replacements
	\begin{align}
	n_{12} &\rightarrow n_{12} + (s_{12}+m_{12}^2)\Delta \nonumber\\
	n_{13} &\rightarrow n_{13} + (s_{13}+m_{13}^2)\Delta \nonumber\\
	n_{14} &\rightarrow n_{14} + (s_{14}+m_{14}^2)\Delta,
	\end{align}
for any function $\Delta$. Putting these together we find the kinematic Jacobi sum of numerators transforms as
	\begin{equation}
	n_{12}+n_{13}+n_{14}\rightarrow n_{12}+n_{13}+n_{14} + \left(m_{12}^2+m_{13}^2+m_{14}^2-m_1^2-m_2^2-m_3^2-m_4^2\right)\Delta.
	\end{equation}
If the spectral condition is not satisfied then we can always find a generalized gauge in which the numerators satisfy color-kinematics duality by using,
\begin{align}
	\Delta&=-\frac{n_{12}+n_{13}+n_{14}}{\left(m_{12}^2+m_{13}^2+m_{14}^2-m_1^2-m_2^2-m_3^2-m_4^2\right)}.
\end{align}

If the spectral condition is satisfied, however, then there is no choice of $\Delta$ that can construct numerators that satisfy the kinematic Jacobi relations from ones that do not. Hence the existence of kinematic Jacobi-satisfying numerators is a non-trivial constraint on the space of BCJ-like models, equivalent to imposing the massive fundamental BCJ relations. 

At 4-point, we saw that there is a well-chosen BCJ basis in which the KLT kernel is polynomial, and therefore together with the discussion in Section \ref{sec:spurious}, the resulting formula defines an amplitude with only physical singularitites. The BCJ version of this statement is that if the spectral condition is satisfied, and there exist color-kinematics duality satisfying numerators, then the BCJ double-copy is free of spurious singularities.

It is clear that a model with a uniform mass spectrum like massive Yang-Mills could only satisfy the 4-point spectral condition if all of the states have zero mass. For more complicated models, with states of multiple masses, the constraints are very restrictive. We will now illustrate these constraints with a few examples.

\subsubsection*{Example 1: Compton Scattering}

Consider a model such as Yang-Mills minimally coupled to a complex adjoint scalar with mass $m\neq 0$. There are three factorization channels contributing to the Compton amplitude $g+\phi \rightarrow g + \phi$:
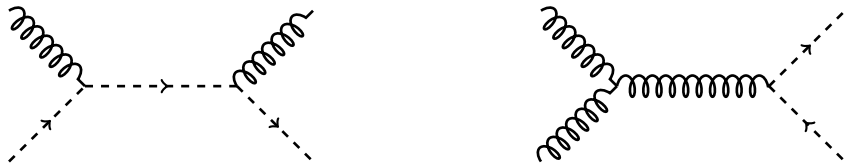
	\begin{center}
	\begin{tikzpicture}[scale=1, line width=1 pt]
	\begin{scope}[shift={(0,0)}]
	\draw[gluon] (-1,1)--(0,0);
	\draw[scalar] (-1,-1)--(0,0);
	\draw[scalar] (0,0)--(2,0);
	\draw[gluon] (2,0)--(3,1);
	\draw[scalar] (2,0)--(3,-1);
	\end{scope}
	\begin{scope}[shift={(7,0)}]
	\draw[gluon] (-1,1)--(0,0);
	\draw[gluon] (-1,-1)--(0,0);
	\draw[gluon] (0,0)--(2,0);
	\draw[scalar] (2,0)--(3,1);
	\draw[scalarbar] (2,0)--(3,-1);
	\end{scope}
	\end{tikzpicture}
\end{center}
The first diagram contributes twice, corresponding to exchanging the labels on the gluons. Here the spectral condition is satisfied since for the external states
\begin{align}
	m_1^2+m_2^2+m_3^2+m_4^2=2m^2,
\end{align}
while for the internal states
\begin{align}
	m_{12}^2+m_{13}^2+m_{14}^2=2m^2.
\end{align}
We must keep in mind that the spectral condition is only a conjectured \textit{necessary} condition for the existence of a local double-copy, not a sufficient one. For a theory to produce a local double-copy, it must also satisfy the BCJ relations. The fact that a sensible double-copy of Compton scattering amplitudes can be defined only if the theory satisfies the massive BCJ relations \eqref{eq:massiveBCJ4} was first observed in \cite{Johansson:2019dnu}. 

Explicitly the color-ordered amplitudes \cite{Naculich:2014naa}
\begin{align}
    &\mathcal{A}_4\left[1_\phi,2_g^+,3_g^-,4_{\overline{\phi}}\right] = -\frac{\langle 3|p_1|2]^2}{s_{23}(s_{12}+m^2)} \nonumber\\
    &\mathcal{A}_4\left[1_\phi,3_g^-,2_g^+,4_{\overline{\phi}}\right] = -\frac{\langle 3|p_1|2]^2}{s_{23}(s_{13}+m^2)} ,
\end{align}
satisfy the massive BCJ relation (\ref{eq:massiveBCJ4}). According to our conjecture the double-copy and the massless limit should commute in such a case. Indeed, taking the massive double-copy and then the massless limit
\begin{equation}
    \mathcal{M}^{m\neq 0}_4\left(1_\phi,2_h^+,3_h^-,4_{\overline{\phi}}\right) = \frac{\langle 3|p_1|2]^4}{(s_{12}+m^2)(s_{13}+m^2)} \xrightarrow[]{m=0} \frac{\langle 3|p_1|2]^4}{s_{12}s_{13}},
\end{equation}
compared to taking the massless limit and then the double-copy
\begin{equation}
    s_{14}\mathcal{A}^{m=0}_4\left[1_\phi,2_g^+,3_g^-,4_{\overline{\phi}}\right]\mathcal{A}^{m=0}_4\left[1_\phi,3_g^-,2_g^+,4_\phi\right] = \frac{\langle 3|p_1|2]^4}{s_{12}s_{13}},
\end{equation}
gives the same result.

\subsubsection*{Example 2: Bhabha Scattering}

In the same model as the previous example we can consider Bhabha scattering $\phi + \overline{\phi} \rightarrow \phi +\overline{\phi}$ which has two contributing factorization channels related by relabelling:
	
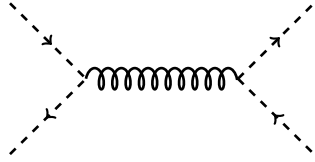
\begin{center}
	\begin{tikzpicture}[scale=1, line width=1 pt]
	\draw[scalar] (-1,1)--(0,0);
	\draw[scalarbar] (-1,-1)--(0,0);
	\draw[gluon] (0,0)--(2,0);
	\draw[scalar] (2,0)--(3,1);
	\draw[scalarbar] (2,0)--(3,-1);
	\end{tikzpicture}
\end{center}
Here the spectral condition is not satisfied since for the external states
\begin{align}
	m_1^2+m_2^2+m_3^2+m_4^2=4m^2,
\end{align}
while for the internal states
\begin{align}
	m_{12}^2+m_{13}^2+m_{14}^2=0.
\end{align}
Since the spectral condition is not satisfied, there are no associated fundamental BCJ conditions. Similar to the 4-point massive Yang-Mills calculation, we can find color-kinematics duality satisfying numerators and take a massive double-copy, but such an amplitude should not have a smooth $m\rightarrow 0$ limit. It is instructive to see this explicitly. We begin with the tree-amplitude calculated using ordinary Feynman rules for a minimally coupled scalar
\begin{equation}
    \mathcal{A}_4\left(1_\phi^{a_1},2_{\overline{\phi}}^{a_2},3_\phi^{a_3},4_{\overline{\phi}}^{a_4}\right) = c_{12} \frac{s_{13}-s_{14}}{s_{12}}+c_{14}\frac{s_{12}-s_{13}}{s_{14}}.
\end{equation}
The corresponding BCJ numerators,
\begin{align}
    n_{12}&=s_{13}-s_{14}\nonumber\\
    n_{13}&=0 \nonumber\\
    n_{14}&=s_{12}-s_{13},
\end{align}
do not satisfy the kinematic Jacobi relation. We can construct numerators which do, however, by making a generalized gauge transformation
\begin{align}
     \hat{n}_{12}&=s_{13}-s_{14}+\frac{1}{4m^2}s_{12}(s_{12}-s_{14})\nonumber\\
    \hat{n}_{13}&=\frac{1}{4m^2}s_{13}(s_{12}-s_{14}) \nonumber\\
    \hat{n}_{14}&=s_{12}-s_{13}+\frac{1}{4m^2}s_{14}(s_{12}-s_{14}).
\end{align}
Forming the massive BCJ double-copy, we find
\begin{align}
    \mathcal{M}_4\left(1_\phi,2_{\overline{\phi}},3_\phi,4_{\overline{\phi}}\right) =  &\frac{(s_{13}-s_{14})^2}{s_{12}}+\frac{(s_{12}-s_{13})^2}{s_{14}}\nonumber\\
    &+4m^2+4 s_{12}+2 s_{13} +\frac{1}{4m^2}\left(4s_{12}^2+4s_{12} s_{13}+s_{13}^2\right).
\end{align}
While this is a perfectly physical scattering amplitude, the massive double-copy has generated a contact contribution corresponding to a local operator of the form $\frac{1}{m^2 M_p^2}(\partial \phi)^4$, which diverges as $m\rightarrow 0$. 

\subsubsection*{Example 3: Kaluza-Klein Theory}

An important class of examples arises from the dimensional reduction of the massless KLT relations in higher dimensions, some of which have already been discussed in \cite{Chiodaroli:2015rdg,Chiodaroli:2017ehv,Chiodaroli:2018dbu,bautista2019double,Bern_2019}. This has the effect of generating a Kaluza-Klein tower of states and vertices that conserve Kaluza-Klein number.  This conservation law manifests as a conservation of mass at each vertex. For concreteness, consider a $d=5$ scalar model compactified on $\mathds{R}^4\times S^1$, and take for example the scattering process $1+2\to 3+4$, where all of the external states are right-moving ($p^4_i=+m_i$) states. At the vertices the masses satisfy the sum rules
\begin{align}
	m_1+m_2&=m_{12}\nonumber\\
	m_1-m_3&=m_{13}\nonumber\\
	m_1-m_4&=m_{14}\nonumber\\
	m_1+m_2&=m_3+m_4.
\end{align}	
In this case as well, the spectral condition holds with no further constraints,
\begin{align}
	\Rightarrow m_{12}^2+m_{13}^2+m_{14}^2&=3m_1^2+m_2^2+m_3^2+m_4^2+2m_1 m_2-2m_1 m_4- 2 m_1 m_3\nonumber\\
	&=3m_1^2+m_2^2+m_3^2+m_4^2-2m_1^2\nonumber\\
	&=m_1^2+m_2^2+m_3^2+m_4^2.
\end{align}
Thus any theory that arises as a dimensional reduction of a massless BCJ-compatible theory will automatically satisfy the spectral condition and thus it will give a local double-copy. Such a model gives a \textit{complete} example, for which every scattering amplitude satisfies the spectral constraints, and moreover, if the higher-dimensional model satisfies the massless BCJ relations then so too will the lower-dimensional Kaluza-Klein model. We leave as future work the problem of determining if there are additional complete examples which are not obtained by dimensional reduction. 

\subsection{5-point Spectral Conditions}
Locality places the strongest constraints on the massive double-copy. As was exemplified in Section \ref{quintic}, demanding the existence of color-kinematics duality satisfying 5-point numerators is not a strong enough condition to ensure locality of double-copied 5-point amplitudes. A natural question is what conditions need to be satisfied at 5-point in order for the resulting double-copied amplitude to be local. 

We set the calculation up in a manner similar to the 4-point case. We assume the existence of a BCJ representation and allow for general external and intermediate masses, $m_i$ and $m_{ij}$ respectively. Here the masses $m_{ij}$ are exchanged on the $ij$ 2-particle channel. We can then write down a bi-adjoint scalar matrix \eqref{eq:biaj5pt} where each propagator $s_{ij}+m^2$ is now replaced by $s_{ij}+m_{ij}^2$.

We know that 5-point amplitudes need to factorize on 2-particle channels to give 4-point amplitudes. At 4-point, we saw that locality is only ensured by requiring that the matrix of bi-adjoint scalar amplitudes is singular. This is achieved via the so-called spectral condition \eqref{eq:speccond}. On demanding that this condition is satisfied on every possible 4-point amplitude that could result on a factorization channel, we come up with the following set of conditions,
\begin{align}
	m_{ij}^2+m_{ik}^2+m_{jk}^2=m_i^2+m_j^2+m_k^2+m_{pq}^2
\end{align}
for each triplet $i$, $j$, $k$ and where $p$, $q$ are the leftover elements in $\{1,2,3,4,5\}$. There are $^5C_3=10$ such relations, but they are not all independent. We can reduce them to 5 independent conditions,
\begin{align}
\label{eq:speccond5}
	m_{15}^2&=2 m_1^2 - m_{12}^2 - m_{13}^2 - m_{14}^2 + m_2^2 + m_3^2 + m_4^2 + m_5^2\nonumber\\
	m_{25}^2&=m_1^2 - m_{12}^2 + 2 m_2^2 - m_{23}^2 - m_{24}^2 + m_3^2 + m_4^2 + m_5^2\nonumber\\
	m_{34}^2&=2 m_1^2 - m_{12}^2 - m_{13}^2 - m_{14}^2 + 2 m_2^2 - m_{23}^2 - m_{24}^2 + 2 m_3^2 + 
	2 m_4^2 + m_5^2\nonumber\\
	m_{35}^2&=-m_1^2 + m_{12}^2 + m_{14}^2 - m_2^2 + m_{24}^2 - m_4^2\nonumber\\
	m_{45}^2&=-m_1^2 + m_{12}^2 + m_{13}^2 - m_2^2 + m_{23}^2 - m_3^2.
\end{align}

We will refer to these as the \textit{5-point spectral conditions}. These conditions indeed make the bi-adjoint scalar matrix singular. Further, they reduce the rank of the $(n-2)!\times(n-2)!=6\times6$ matrix from full-rank to minimal rank, $(n-3)!=2$. 

As we show in Appendix \ref{app:BCJnull}, the null vectors of the bi-adjoint scalar matrix give us the 5-point massive BCJ relations,
\begin{align}
	\mathcal{A}_5[13452]=&\left(-\frac{m_{12}^2+s_{12}}{m_{34}^2+s_{34}}+\frac{m_{35}^2+s_{35}}{m_{34}^2+s_{34}}\right)\mathcal{A}_5[13542]+\left(\frac{m_{14}^2+s_{14}}{m_{34}^2+s_{34}}\right)\mathcal{A}_5[13524]\,,\\
	\mathcal{A}_5[13425]=&\left(\frac{\left(m_{12}^2+s_{12}\right) \left(m_{45}^2+s_{45}\right)}{\left(m_{15}^2+s_{15}\right) \left(m_{34}^2+s_{34}\right)}\right)\mathcal{A}_5[13542]\nonumber\\
	&\hspace{2.5cm}+\left(\frac{m_{14}^2+s_{14}}{m_{15}^2+s_{15}}-\frac{\left(m_{12}^2+s_{12}\right) \left(m_{14}^2+s_{14}\right)}{\left(m_{15}^2+s_{15}\right)\left(m_{34}^2+s_{34}\right)}\right)\mathcal{A}_5[13524]\,,\\
	\mathcal{A}_5[13245]=&\left(\frac{m_{12}^2+s_{12}}{m_{15}^2+s_{15}}-\frac{\left(m_{12}^2+s_{12}\right) \left(m_{14}^2+s_{14}\right)}{\left(m_{15}^2+s_{15}\right)\left(m_{23}^2+s_{23}\right)}\right)\mathcal{A}_5[13542]\nonumber\\
	&\hspace{5cm}+\left(\frac{\left(m_{14}^2+s_{14}\right) \left(m_{25}^2+s_{25}\right)}{\left(m_{15}^2+s_{15}\right) \left(m_{23}^2+s_{23}\right)}\right)\mathcal{A}_5[13524]\,,\\
	\label{eq:massBCJ1}
	\mathcal{A}_5[13254]=&\left(-\frac{m_{12}^2+s_{12}}{m_{23}^2+s_{23}}\right)\mathcal{A}_5[13542]+\left(-\frac{m_{12}^2+s_{12}}{m_{23}^2+s_{23}}-\frac{m_{24}^2+s_{24}}{m_{23}^2+s_{23}}\right)\mathcal{A}_5[13524]\,,
\end{align}
with the understanding that $m_{15}$, $m_{25}$, $m_{34}$, $m_{35}$ and $m_{45}$ are given by the spectral conditions \eqref{eq:speccond5}.

Choosing any $2\times 2$ submatrix $\mathcal{A}^{\phi^3}[\alpha|\beta]$ of the bi-adjoint scalar matrix is now invertible and can be used to define a local double-copy. For example,
\begin{align}
\label{eq:420}
	\mathcal{A}^{A\otimes B}_5(12345)&=\sum_{\alpha,\beta=[13542],[13524]}\mathcal{A}^A_5[\alpha]\  \mathcal{A}^{\phi^3}[\alpha|\beta]^{-1}\ \mathcal{A}^A_5[\beta],
\end{align}
where
\begin{align}
	\mathcal{A}^{\phi^3}[\alpha|\beta]&=\begin{pmatrix}
	\frac{1}{D_1}+\frac{1}{D_{12}}+\frac{1}{D_2}+\frac{1}{D_6}+\frac{1}{D_9}  &  -\frac{1}{D_{12}}-\frac{1}{D_9} \\
	\\
	-\frac{1}{D_{12}}+\frac{1}{D_9} &  \frac{1}{D_{12}}+\frac{1}{D_3}+\frac{1}{D_4}+\frac{1}{D_5}+\frac{1}{D_9}
	\end{pmatrix},
\end{align}
and $D_i$ are as defined in Appendix \ref{app:5ptmatrix}.

To explicitly see that the resulting amplitude is local, we perform the following tests. First, we look at the denominator of the resulting KLT formula,
\begin{align}
	(s_{15} + m_{15}^2) (m_{23}^2 + s_{23}) (s_{34} + m_{34}^2),
\end{align}
again with the understanding that $m_{ij}$ satisfy the spectral conditions \eqref{eq:speccond5} and note that the KLT formula only has poles in physical locations.

Second, we must check that $\mathcal{A}^{A\otimes B}_5(12345)$ factorizes correctly on all poles. Let us look at an example. Consider the pole $s_{23}\to -m_{23}^2$,
\begin{align}
	\underset{s_{23}=-m_{23}^2}{\text{Res}}\mathcal{A}^{A\otimes B}_5(12345)=&\frac{\left(m_{14}^2+s_{14}\right) \left(m_{12}^2+m_{13}^2+s_{12}+s_{13}\right)}{\left(s_{15}+m_{15}^2\right)}\bigg[\mathcal{A}_5[13542]\left(m_{12}^2+s_{12}\right)\nonumber\\
	&\hspace{3.5cm}+\mathcal{A}_5[13524] \left(m_{12}^2+m_{24}^2+s_{12}+s_{24}\right)\bigg]^2.
\end{align}

The massive BCJ relation \eqref{eq:massBCJ1} tells us that the expression in the square brackets is $\mathcal{A}_5[13254]$ which factorizes into $\mathcal{A}_3[32(-P_{23})]\times \mathcal{A}_4[P_{23}541]$ on the pole to give
\begin{align}
\underset{s_{23}=-m_{23}^2}{\text{Res}}\hspace{-1mm}\mathcal{A}^{A\otimes B}_5(12345)=&\frac{\left(m_{14}^2+s_{14}\right) \left(m_{12}^2+m_{13}^2+s_{12}+s_{13}\right)}{\left(s_{15}+m_{15}^2\right)}\left(\mathcal{A}_3[32(-P_{23})]\ \mathcal{A}_4[P_{23}541]\right)^2\nonumber\\
=&\mathcal{A}^{A\otimes B}_3(32(-P_{23}))\times \mathcal{A}^{A\otimes B}_4(P_{23}541)\,,
\end{align}
where we have used the 4-point KLT formula in the last step. Thus the amplitude factorizes correctly on the $s_{23}=-m_{23}^2$ pole. 

One can proceed in a similar manner (either with or without the help of massive BCJ relations) to determine that the 5-point KLT formula \eqref{eq:420} factorizes correctly on all poles. Thus, \textit{given a theory that satisfies the 5-point spectral conditions, the KLT formula constructs local amplitudes}, giving us a sensible definition of the 5-point double-copy.

\subsection{Non-minimal Rank}
There is a new possibility that arises at higher-point which is not present at 4-point. This is the ability to reduce the rank of a bi-adjoint scalar matrix from full-rank $(n-2)!$ not to minimal rank $(n-3)!$, but somewhere in between $(n-2)!$ and $(n-3)!$. Since this too makes the $(n-2)!\times (n-2)!$ matrix singular, one might imagine this to be an alternate approach to the massive double-copy that does not require all four BCJ relations to hold. Indeed such a procedure \textit{does not} give rise to local amplitudes. Let us understand how this works at 5-point.

By imposing all-but-one of the spectral conditions \eqref{eq:speccond5}, the rank of the 5-point bi-adjoint scalar matrix reduces from 6 to 4, rather than minimal rank 2. For example, let us choose not to impose the spectral condition on $m_{34}^2$. Since the resulting expressions are difficult to manipulate analytically, we proceed in a particular kinematic configuration where all-but-one (let us say $s_{12}$) independent Mandelstam variables are fixed.

We can now check the behaviour of the double-copied amplitude as we approach the pole $s_{12}=-m_{12}^2$. We want the double-copied amplitude to factorize as,
\begin{align}
	\underset{s_{12}=-m_{12}^2}{\text{Res}}\mathcal{A}^{A\otimes B}_5(12345)=&\mathcal{A}^{A\otimes B}_3(12(-P_{12}))\times \mathcal{A}^{A\otimes B}_4(P_{12}345).
\end{align}
We find that this condition is not met unless,
\begin{align}
	m_{34}^2&=2 m_1^2 - m_{12}^2 - m_{13}^2 - m_{14}^2 + 2 m_2^2 - m_{23}^2 - m_{24}^2 + 2 m_3^2 + 
	2 m_4^2 + m_5^2,
\end{align}
which is exactly the spectral condition that we left out. Thus, by not imposing all of the BCJ relations, we do not construct local amplitudes.

This supports our conjecture: only by imposing all BCJ relations, i.e. reducing the bi-adjoint scalar matrix to minimal rank, can we construct local amplitudes via the KLT formula.

\section{Discussion}
\label{sec:conclusions}
The proposition of a KLT construction for the double-copy of massive particles opens up many areas of exploration and application. In Section \ref{sec:mYM2}, we see that the double-copy of massive Yang-Mills is ill-defined due to the presence of spurious singularities in the would-be double-copied 5-point amplitude. It is still left as an open question whether or not this construction can be salvaged. For example, can we add 5-point operators or new degrees of freedom to the massive Yang-Mills EFT to construct a local double-copy?

Another interesting question is what happens when the bi-adjoint Higgs model presented in Section \ref{sec:mYM2} is double copied with itself. It has been shown that the high energy behaviour of a theory of $\Lambda_3$ massive gravity cannot be improved by introducing vector or scalar interactions \cite{Bonifacio:2019mgk}. Therefore, we expect the double-copy of the bi-adjoint Higgs model to fail. A better understanding of the precise nature of this failure would be interesting.

An important assumption that lead to the derivation of the mass spectral conditions presented in Section \ref{sec:diffmass} was that a unique mass is exchanged in each factorization channel. We know that a massless KLT formula can be constructed that allows for the exchange of particles of multiple masses on each channel \cite{Mizera:2016jhj}. It would be interesting to see how this construction generalizes to the case of massive external particles and more general spectra.

 In addition, we would like to better understand the landscape of theories that produce a local double-copy. We saw examples of dimensionally reduced BCJ-compatible theories in which the Kaluza-Klein tower of massive states and interactions between them manifestly satisfy the spectral condition and hence result in local double-copied amplitudes. We would like to understand whether there are double-copy-compatible theories that do not result from a dimensional reduction.

Finally, in Section \ref{sec:diffmass}, we saw that spurious singularities are removed \textit{if} the spectral conditions and massive BCJ relations are satisfied. However, we know that massive bi-adjoint scalar theory trivially provides an explicit counter-example to making the converse statement, since it will produce a local, massive double-copy even if the spectral conditions are not satisfied. It is therefore an interesting open problem to determine if there exist further, non-trivial, examples of massive models which double-copy to physical scattering amplitudes but do not satisfy the spectral condition. One pathway to such a construction would be to try and find a model which admits a local, off-shell representation of the kinematic algebra, similar to \cite{Cheung:2016prv,Monteiro:2011pc}. Since the numerators of such a model are local by construction, it is clear from the BCJ form of the double-copy that no spurious poles can be generated. Even more interestingly, given such a set of local, kinematic Jacobi satisfying numerators, we can always form a heterotic double-copy with the numerators of a generic, spectral condition violating, massive model. Since the result does not depend on the generalized gauge used for the numerators of the latter, they can always be taken to be the local representation given by Feynman rules, and so even in this case, we see that no spurious poles can be generated. We see then that constructing even a single example of a model with a local, off-shell representation of the kinematic algebra, is sufficient to generate an infinite number of examples of healthy, massive double-copies. We leave this and similar investigations to future work.

\subsection*{Acknowledgement}
We would like to thank James Bonifacio, Clifford Cheung, Henriette Elvang, Kurt Hinterbichler, Henrik Johansson and Andrew Tolley for useful discussions. The work of CRTJ was supported by a Rackham Predoctoral Fellowship from the University of Michigan. The work of SP was supported in part by the US Department of Energy under Grant No. DE-SC0007859. We would also like to acknowledge the TASI 2019 summer school, which brought us together and played host to preliminary discussions.

\begin{figure}[h]
\begin{center}
 \begin{tikzpicture}[scale=0.9, line width=1 pt]
\begin{scope}[shift={(0,0)}]
\begin{scope}[shift={(0,0)}]
  \draw[gluon] (0,1)--(0,-1);
  \draw[gluon] (0,0)--(2,0);
  \draw[gluon] (2,1)--(2,-1);
  \draw[gluon] (1,0)--(1,1);
  \node at (0,-1.3) {1};
  \node at (0,1.3) {2};
  \node at (1,1.3) {3};
  \node at (2,1.3) {4};
  \node at (2,-1.3) {5};
  \node at (1,-1.3) {(1)};
\end{scope}
\begin{scope}[shift={(3.5,0)}]
  \draw[gluon] (0,1)--(0,-1);
  \draw[gluon] (0,0)--(2,0);
  \draw[gluon] (2,1)--(2,-1);
  \draw[gluon] (1,0)--(1,1);
  \node at (0,-1.3) {1};
  \node at (0,1.3) {3};
  \node at (1,1.3) {2};
  \node at (2,1.3) {4};
  \node at (2,-1.3) {5};
  \node at (1,-1.3) {(2)};
\end{scope}
\begin{scope}[shift={(7,0)}]
  \draw[gluon] (0,1)--(0,-1);
  \draw[gluon] (0,0)--(2,0);
  \draw[gluon] (2,1)--(2,-1);
  \draw[gluon] (1,0)--(1,1);
  \node at (0,-1.3) {1};
  \node at (0,1.3) {3};
  \node at (1,1.3) {4};
  \node at (2,1.3) {2};
  \node at (2,-1.3) {5};
  \node at (1,-1.3) {(3)};
\end{scope}
\begin{scope}[shift={(10.5,0)}]
  \draw[gluon] (0,1)--(0,-1);
  \draw[gluon] (0,0)--(2,0);
  \draw[gluon] (2,1)--(2,-1);
  \draw[gluon] (1,0)--(1,1);
  \node at (0,-1.3) {1};
  \node at (0,1.3) {4};
  \node at (1,1.3) {3};
  \node at (2,1.3) {2};
  \node at (2,-1.3) {5};
  \node at (1,-1.3) {(4)};
\end{scope}
\begin{scope}[shift={(14,0)}]
  \draw[gluon] (0,1)--(0,-1);
  \draw[gluon] (0,0)--(2,0);
  \draw[gluon] (2,1)--(2,-1);
  \draw[gluon] (1,0)--(1,1);
  \node at (0,-1.3) {1};
  \node at (0,1.3) {4};
  \node at (1,1.3) {2};
  \node at (2,1.3) {3};
  \node at (2,-1.3) {5};
  \node at (1,-1.3) {(5)};
\end{scope}
\end{scope}
\begin{scope}[shift={(0,-3.5)}]
\begin{scope}[shift={(0,0)}]
  \draw[gluon] (0,1)--(0,-1);
  \draw[gluon] (0,0)--(2,0);
  \draw[gluon] (2,1)--(2,-1);
  \draw[gluon] (1,0)--(1,1);
  \node at (0,-1.3) {1};
  \node at (0,1.3) {2};
  \node at (1,1.3) {4};
  \node at (2,1.3) {3};
  \node at (2,-1.3) {5};
  \node at (1,-1.3) {(6)};
\end{scope}
\begin{scope}[shift={(3.5,0)}]
  \draw[gluon] (0,1)--(0,-1);
  \draw[gluon] (0,0)--(2,0);
  \draw[gluon] (2,1)--(2,-1);
  \draw[gluon] (1,0)--(1,1);
  \node at (0,-1.3) {1};
  \node at (0,1.3) {2};
  \node at (1,1.3) {5};
  \node at (2,1.3) {3};
  \node at (2,-1.3) {4};
  \node at (1,-1.3) {(7)};
\end{scope}
\begin{scope}[shift={(7,0)}]
  \draw[gluon] (0,1)--(0,-1);
  \draw[gluon] (0,0)--(2,0);
  \draw[gluon] (2,1)--(2,-1);
  \draw[gluon] (1,0)--(1,1);
  \node at (0,-1.3) {2};
  \node at (0,1.3) {3};
  \node at (1,1.3) {1};
  \node at (2,1.3) {4};
  \node at (2,-1.3) {5};
  \node at (1,-1.3) {(8)};
\end{scope}
\begin{scope}[shift={(10.5,0)}]
  \draw[gluon] (0,1)--(0,-1);
  \draw[gluon] (0,0)--(2,0);
  \draw[gluon] (2,1)--(2,-1);
  \draw[gluon] (1,0)--(1,1);
  \node at (0,-1.3) {1};
  \node at (0,1.3) {3};
  \node at (1,1.3) {5};
  \node at (2,1.3) {2};
  \node at (2,-1.3) {4};
  \node at (1,-1.3) {(9)};
\end{scope}
\begin{scope}[shift={(14,0)}]
  \draw[gluon] (0,1)--(0,-1);
  \draw[gluon] (0,0)--(2,0);
  \draw[gluon] (2,1)--(2,-1);
  \draw[gluon] (1,0)--(1,1);
  \node at (0,-1.3) {3};
  \node at (0,1.3) {4};
  \node at (1,1.3) {1};
  \node at (2,1.3) {2};
  \node at (2,-1.3) {5};
  \node at (1,-1.3) {(10)};
\end{scope}
\end{scope}
\begin{scope}[shift={(0,-7)}]
\begin{scope}[shift={(0,0)}]
  \draw[gluon] (0,1)--(0,-1);
  \draw[gluon] (0,0)--(2,0);
  \draw[gluon] (2,1)--(2,-1);
  \draw[gluon] (1,0)--(1,1);
  \node at (0,-1.3) {2};
  \node at (0,1.3) {3};
  \node at (1,1.3) {5};
  \node at (2,1.3) {1};
  \node at (2,-1.3) {4};
  \node at (1,-1.3) {(11)};
\end{scope}
\begin{scope}[shift={(3.5,0)}]
  \draw[gluon] (0,1)--(0,-1);
  \draw[gluon] (0,0)--(2,0);
  \draw[gluon] (2,1)--(2,-1);
  \draw[gluon] (1,0)--(1,1);
  \node at (0,-1.3) {2};
  \node at (0,1.3) {4};
  \node at (1,1.3) {1};
  \node at (2,1.3) {3};
  \node at (2,-1.3) {5};
  \node at (1,-1.3) {(12)};
\end{scope}
\begin{scope}[shift={(7,0)}]
  \draw[gluon] (0,1)--(0,-1);
  \draw[gluon] (0,0)--(2,0);
  \draw[gluon] (2,1)--(2,-1);
  \draw[gluon] (1,0)--(1,1);
  \node at (0,-1.3) {3};
  \node at (0,1.3) {4};
  \node at (1,1.3) {2};
  \node at (2,1.3) {1};
  \node at (2,-1.3) {5};
  \node at (1,-1.3) {(13)};
\end{scope}
\begin{scope}[shift={(10.5,0)}]
  \draw[gluon] (0,1)--(0,-1);
  \draw[gluon] (0,0)--(2,0);
  \draw[gluon] (2,1)--(2,-1);
  \draw[gluon] (1,0)--(1,1);
  \node at (0,-1.3) {2};
  \node at (0,1.3) {3};
  \node at (1,1.3) {4};
  \node at (2,1.3) {1};
  \node at (2,-1.3) {5};
  \node at (1,-1.3) {(14)};
\end{scope}
\begin{scope}[shift={(14,0)}]
  \draw[gluon] (0,1)--(0,-1);
  \draw[gluon] (0,0)--(2,0);
  \draw[gluon] (2,1)--(2,-1);
  \draw[gluon] (1,0)--(1,1);
  \node at (0,-1.3) {2};
  \node at (0,1.3) {4};
  \node at (1,1.3) {3};
  \node at (2,1.3) {1};
  \node at (2,-1.3) {5};
  \node at (1,-1.3) {(15)};
\end{scope}
\end{scope}
\end{tikzpicture}
 \caption{Color-dressed tree-level 5-point amplitude organized using graphs with only cubic vertices. \label{5ptYM}}
 \end{center}
 \end{figure}
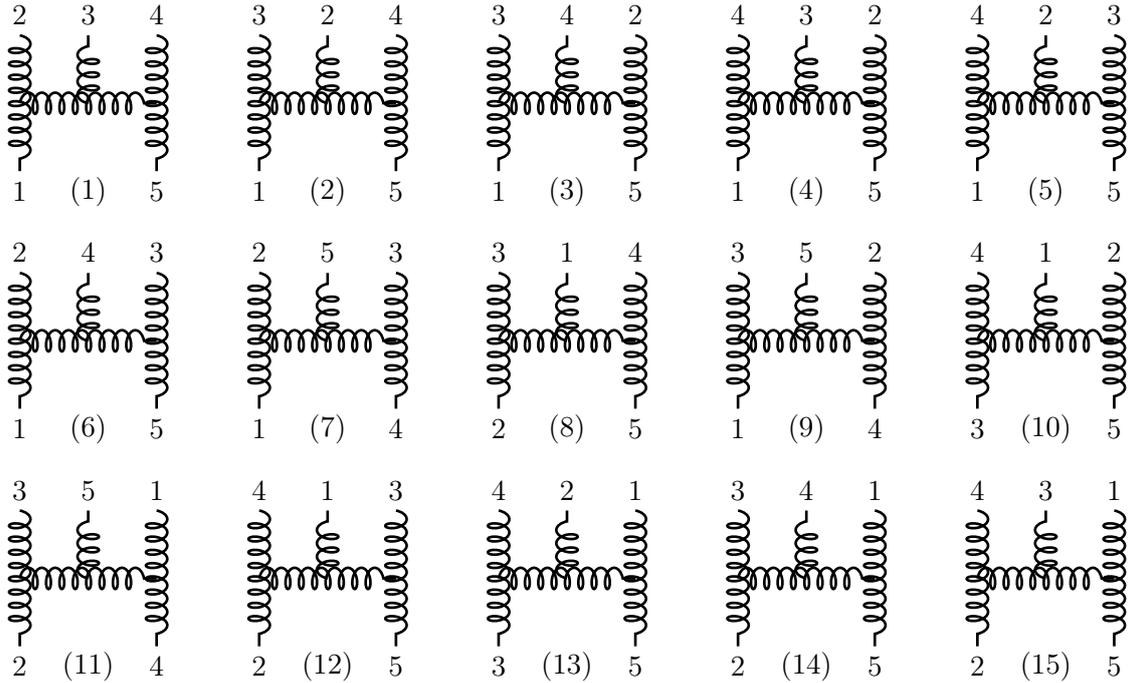

\appendix
\begin{landscape}
\section{Matrix of 5-point Bi-adjoint Scalar
Amplitudes}
\label{app:5ptmatrix}
The numerators are given by $\begin{pmatrix}
n_1\\ n_2\\ n_3\\ n_4\\ n_5\\ n_6\end{pmatrix}=
 M_n^{-1} \begin{pmatrix}
\mathcal{A}_5[13542]\\ \mathcal{A}_5[13524]\\ \mathcal{A}_5[13254]\\ \mathcal{A}_5[13245]\\ \mathcal{A}_5[13425]\\ \mathcal{A}_5[13452]\end{pmatrix}$ where $M_n$ is given by:
\begin{align}
&M_n=\nonumber\\
&\nonumber\\
&\begin{pmatrix}
\frac{1}{D_1}  & \quad \frac{1}{D_2}+\frac{1}{D_9} & \quad -\frac{1}{D_9} & \quad 0 & \quad -\frac{1}{D_{12}}& \quad\frac{1}{D_{12}}+\frac{1}{D_6} \\
\\
0  & \quad -\frac{1}{D_9}&\quad \frac{1}{D_3}+\frac{1}{D_9} & \quad \frac{1}{D_4} & \quad \frac{1}{D_{12}}+\frac{1}{D_5} & \quad -\frac{1}{D_{12}} \\
\\
\frac{1}{D_8}  & \quad -\frac{1}{D_2}-\frac{1}{D_8} & \quad -\frac{1}{D_3} & \quad -\frac{1}{D_{11}}- \frac{1}{D_4} & \quad \frac{1}{D_{11}}& 0 \\
\\
-\frac{1}{D_{14}}-\frac{1}{D_8}  & \quad \frac{1}{D_{14}}+\frac{1}{D_{15}}+\frac{1}{D_2}+\frac{1}{D_8}+\frac{1}{D_9} & \quad -\frac{1}{D_{15}}-\frac{1}{D_9} & \quad -\frac{1}{D_{14}} & \quad \frac{1}{D_{14}}+\frac{1}{D_{15}}& -\frac{1}{D_{15}} \\
\\
-\frac{1}{D_{13}}  & \quad -\frac{1}{D_{15}}-\frac{1}{D_9} & \quad \frac{1}{D_{10}}+\frac{1}{D_{13}}+\frac{1}{D_{15}}+\frac{1}{D_3}+\frac{1}{D_9} & \quad -\frac{1}{D_{10}}-\frac{1}{D_{13}} & \quad -\frac{1}{D_{15}}& \quad\frac{1}{D_{13}}+\frac{1}{D_{15}} \\
\\
-\frac{1}{D_1}-\frac{1}{D_7}  & \quad -\frac{1}{D_2}&\quad-\frac{1}{D_{10}}-\frac{1}{D_3} & \quad \frac{1}{D_{10}} & \quad 0 & \quad -\frac{1}{D_7} \\
\end{pmatrix},
\end{align}
where $1/D_i$ corresponds to the propagators used in \cite{Bern:2019prr} for the 5-point trivalent graphs shown in Figure \ref{5ptYM}. 
The 5-point amplitude can be calculated from the bi-adjoint scalar matrix as
\begin{equation}
\mathcal{A}_5^{A\otimes B}(1,2,3,4,5)=\begin{pmatrix}
\mathcal{A}_5^{A}[13542]& \mathcal{A}_5^{A}[13524]& \mathcal{A}_5^{A}[13254]& \mathcal{A}_5^{A}[13245]& \mathcal{A}_5^{A}[13425]& \mathcal{A}_5^{A}[13452]\end{pmatrix}
  \left(\mathcal{A}_5^{\phi^3}\right)^{-1}[\alpha|\beta] \begin{pmatrix}
\mathcal{A}_5^{B}[13542]\\ \mathcal{A}_5^{B}[13524]\\ \mathcal{A}_5^{B}[13254]\\ \mathcal{A}_5^{B}[13245]\\ \mathcal{A}_5^{B}[13425]\\ \mathcal{A}_5^{B}[13452]\end{pmatrix}\,,
\end{equation} where the bi-adjoint scalar matrix is given by: 

\begin{align}
\label{eq:biaj5pt}
&\mathcal{A}^{\phi^3}_5[\alpha|\beta]=\nonumber\\
&\nonumber\\
&{\tiny 
\begin{pmatrix}
\frac{1}{D_1}+\frac{1}{D_{12}}+\frac{1}{D_2}+\frac{1}{D_6}+\frac{1}{D_9}  &  -\frac{1}{D_{12}}-\frac{1}{D_9} &  -\frac{1}{D_2} &  \frac{1}{D_2}+\frac{1}{D_9} &  -\frac{1}{D_9}&  -\frac{1}{D_1}-\frac{1}{D_2} \\
\\
-\frac{1}{D_{12}}-\frac{1}{D_9} &  \frac{1}{D_{12}}+\frac{1}{D_3}+\frac{1}{D_4}+\frac{1}{D_5}+\frac{1}{D_9} & -\frac{1}{D_3}-\frac{1}{D_4} &  -\frac{1}{D_9} &  \frac{1}{D_3}+\frac{1}{D_9} &  -\frac{1}{D_3} \\
\\
-\frac{1}{D_2}  &  -\frac{1}{D_3}-\frac{1}{D_4} &  \frac{1}{D_{11}}+\frac{1}{D_2}+\frac{1}{D_3}+\frac{1}{D_4}+\frac{1}{D_8} &  -\frac{1}{D_2}- \frac{1}{D_8} &  -\frac{1}{D_3}& \frac{1}{D_2}+\frac{1}{D_3} \\
\\
\frac{1}{D_2}+\frac{1}{D_9}  &  -\frac{1}{D_9} &  -\frac{1}{D_2}-\frac{1}{D_8} &  \frac{1}{D_{14}}+\frac{1}{D_{15}}+\frac{1}{D_2}+\frac{1}{D_8}+\frac{1}{D_9} &  -\frac{1}{D_{15}}-\frac{1}{D_9}& -\frac{1}{D_2} \\
\\
-\frac{1}{D_9}  &  \frac{1}{D_3}+\frac{1}{D_9} &  -\frac{1}{D_3} &  -\frac{1}{D_{15}}-\frac{1}{D_9} &  \frac{1}{D_{10}}+\frac{1}{D_{13}}+\frac{1}{D_{15}}+\frac{1}{D_3}+\frac{1}{D_9}&  -\frac{1}{D_{10}}-\frac{1}{D_3} \\
\\
-\frac{1}{D_1}-\frac{1}{D_2}  &  -\frac{1}{D_3}&\frac{1}{D_2}+\frac{1}{D_3} &  -\frac{1}{D_2} &  -\frac{1}{D_{10}}-\frac{1}{D_3} &  \frac{1}{D_1}+\frac{1}{D_{10}}+\frac{1}{D_2}+\frac{1}{D_3}+\frac{1}{D_7}\\
\end{pmatrix}. }
\end{align}
\end{landscape}

\section{Multi-Particle Factorization}
\label{app:factor}
The argument in Section \ref{sec:factor} generalizes straightforwardly to multi-particle factorization. Without loss of generality we will consider factorization on the singularity
\begin{equation}
P^2 = m^2, \;\;\;\; \text{where }P^\mu \equiv p_1^\mu + p_2^\mu + ... + p_{k-1}^\mu + p_k^\mu.
\end{equation}
A double-ordered bi-adjoint scalar amplitude will contain such a singularity only if both its orderings have $\{1,2,...,k\}$ cyclically adjacent. As we did in Section \ref{sec:factor}, we choose a DDM basis for the $n$-point amplitudes in which the \textit{minimal} number of amplitudes with a $P^2$ factorization singularity appear. A natural choice is 
\begin{equation}
\{\mathcal{A}^{\phi^3}_n[1,\alpha,n|1,\beta,n]:\alpha,\beta \in \mathcal{P}\left(2,3,...,n-1\right)\}.
\end{equation}
The subset of these amplitudes which have a $P^2=m^2$ singularity have the form
\begin{equation}
\{\mathcal{A}^{\phi^3}_n[1,\sigma,\rho,n|1,\sigma',\rho',n]:\sigma,\sigma' \in \mathcal{P}\left(2,3,...,k-1,k\right),\; \rho,\rho'\in \mathcal{P}\left(k+1,k+2,...,n-1,n\right)\}.
\end{equation}
Near the singularity such amplitudes have the form
\begin{equation}
\mathcal{A}^{\phi^3}_n[1,\sigma,\rho,n|1,\sigma',\rho',n] = \frac{\mathcal{A}^{\phi^3}_{k+1}[1,\sigma,-P|1,\sigma',-P]\mathcal{A}^{\phi^3}_{k+1}[P,\rho,n|P,\rho',n]}{P^2+m^2} + \mathcal{O}\left((P^2+m^2)^0\right).
\end{equation}
Placing all such amplitudes in the top-left-hand corner of the matrix of biadjoint-scalar amplitudes, we obtain the same result as in Section \ref{sec:spurious}, that only amplitudes of this form are important on the factorization channel when using the block decomposition inverse formula \eqref{eq:invblock}. Here the associated subspaces are indexed by a \textit{pair} of orderings $(\sigma,\rho)$ on the left and $(\sigma',\rho')$ on the right. The required inverse is then given by
\begin{align}
&\left(\mathcal{A}^{\phi^3}_n\right)^{-1}[1,\sigma,\rho,n|1,\sigma',\rho',n] \nonumber\\
&	\hspace{2mm}=(P^2+m^2)\left(\mathcal{A}^{\phi^3}_{k+1}\right)^{-1}[1,\sigma,-P|1,\sigma',-P]\left(\mathcal{A}^{\phi^3}_{k+1}\right)^{-1}[P,\rho,n|P,\rho',n] + \mathcal{O}\left((P^2+m^2)^2\right).
\end{align}
This is an application of a general result for the so-called \textit{Kronecker} product of matrices
\begin{equation}
(P\otimes Q)^{-1} = P^{-1}\otimes Q^{-1}.
\end{equation}
Verifying that this is true is trivial in component form. We label the components as $P_{ik}$ and $Q_{jl}$, the Kronecker product is then defined component-wise as $(P\otimes Q)_{ijkl} \equiv P_{ik}Q_{jl}$. The right-inverse is defined to satisfy
\begin{equation}
\sum_{m,n}(P\otimes Q)_{ijmn}(P\otimes Q)^{-1}_{mnkl} =\delta_{ik}\delta_{jl}. 
\end{equation}
It is straightforward to see that this is satisfied by matrices of the form
\begin{equation}
(P\otimes Q)^{-1}_{mnkl} = (P^{-1})_{mk}(Q^{-1})_{nl},
\end{equation}
and similarly for the left-inverse. Using this result, on the neighborhood of the $P^2=m^2$ pole,
\begin{align*}
&\mathcal{A}_n^{A\otimes B}\left(1,2,...,n\right) \nonumber\\
&= \sum_{\alpha,\beta} \mathcal{A}_n^A[\alpha] \left(\mathcal{A}_n^{\phi^3}\right)^{-1}[\alpha|\beta] \mathcal{A}^B_n[\beta] \nonumber\\
&= \sum_{\sigma,\sigma'} \sum_{\rho\rho'} \frac{1}{(P^2+m^2)^2}\bigg(\mathcal{A}_{k+1}^A[1,\sigma,-P]\mathcal{A}_{n-k+1}^A[P,\rho,n]\times\left(\mathcal{A}^{\phi^3}_n\right)^{-1}[1,\sigma,\rho,n|1,\sigma',\rho',n]\nonumber\\
&\hspace{6cm}\times\mathcal{A}_{k+1}^B[1,\sigma',-P]\mathcal{A}_{n-k+1}^B[P,\rho',n]\bigg)+ \mathcal{O}\left((P^2+m^2)^0\right) \nonumber
\end{align*}
\begin{align}
&= \sum_{\sigma,\sigma'} \sum_{\rho\rho'} \frac{1}{P^2+m^2}\left(\mathcal{A}_{k+1}^A[1,\sigma,-P]\mathcal{A}_{n-k+1}^A[P,\rho,n]\left(\mathcal{A}^{\phi^3}_{k+1}\right)^{-1}[1,\sigma,-P|1,\sigma',-P]\right. \nonumber\\
&\hspace{35mm}\left. \times\left(\mathcal{A}^{\phi^3}_{n-k+1}\right)^{-1}[P,\rho,n|P,\rho',n]\mathcal{A}_{k+1}^B[1,\sigma',-P]\mathcal{A}_{n-k+1}^B[P,\rho',n]\right) \nonumber\\
&\hspace{12cm}+ \mathcal{O}\left((P^2+m^2)^0\right) \nonumber\\
&=  \frac{1}{P^2+m^2}\left(\sum_{\sigma,\sigma'}\mathcal{A}_{k+1}^A[1,\sigma,-P]\left(\mathcal{A}^{\phi^3}_{k+1}\right)^{-1}[1,\sigma,-P|1,\sigma',-P]\mathcal{A}_{k+1}^B[1,\sigma',-P]\right)\nonumber\\
&\hspace{30mm}\times\left(\sum_{\rho\rho'} \mathcal{A}_{n-k+1}^A[P,\rho,n]\left(\mathcal{A}^{\phi^3}_{n-k+1}\right)^{-1}[P,\rho,n|P,\rho',n]\mathcal{A}_{n-k+1}^B[P,\rho',n]\right) \nonumber\\
&\hspace{12cm}+ \mathcal{O}\left((P^2+m^2)^0\right) \nonumber\\
&= \frac{\mathcal{A}_{k+1}^{A\otimes B}\left(1,2,...,k,-P\right) \mathcal{A}_{n-k+1}^{A\otimes B}\left(P,k+1,...,n\right) }{P^2+m^2} + \mathcal{O}\left((P^2+m^2)^0\right).
\end{align}
So we find that the massive KLT formula generates expressions which factor correctly on all singularities. 

\section{Feynman Rules for Massive Yang-Mills}
\label{app:Feyn}
At low multiplicity it is efficient to calculate the scattering amplitudes of massive Yang-Mills (\ref{mYML}) using Feynman rules. The vertex functions are identical to those of standard non-Abelian gauge theory:
\begin{center}
\begin{tikzpicture}[scale=1, line width=1 pt]
    \draw[gluon] (-2,0)--(0,0); 
    \draw[gluon] (0,0)--(1,1.73);
    \draw[gluon] (0,0)--(1,-1.73);
    \draw [->] (-0.5,0.5)--(-1.5,0.5);
    \draw [->] (0.7,0.3)--(1.2,1.066);
    \draw [->] (-0.15,-0.7)--(0.29,-1.566);
    \node at (-2.6,0) {$\mu_1, \; a_1$};
    \node at (1.05,2) {$\mu_2, \; a_2$};
    \node at (1.05,-2) {$\mu_3, \; a_3$};
    \node at (-1,0.8) {$p_1$};
    \node at (1.3,0.5) {$p_2$};
    \node at (-0.3,-1.3) {$p_3$};
    \node at (2.5,0) {$=$};
    \node at (6.5,0.5) {
    \begin{minipage}{7cm}
    \begin{align*}
        gf^{a_1 a_2 a_3} & \left[ g^{\mu_1 \mu_2}\left(p_2^{\mu_3}-p_1^{\mu_3}\right)\right.\\
        &\left. +g^{\mu_2 \mu_3}\left(p_3^{\mu_1}-p_2^{\mu_1}\right)\right.\\
        &\left. +g^{\mu_3 \mu_1}\left(p_1^{\mu_2}-p_3^{\mu_2}\right)\right],
    \end{align*}
    \end{minipage}
    };
\end{tikzpicture}
\end{center}
\vspace{5mm}
\begin{center}
\begin{tikzpicture}[scale=1, line width=1 pt]
    \node at (-3.15,0) {}; 
    \draw[gluon] (-1.5,1.5)--(0,0); 
    \draw[gluon] (-1.5,-1.5)--(0,0); 
    \draw[gluon] (0,0)--(1.5,1.5);
    \draw[gluon] (0,0)--(1.5,-1.5);
    \draw [->] (-0.3,0.8)--(-0.9,1.4);
    \draw [->] (0.8,0.3)--(1.4,0.9);
    \draw [->] (0.3,-0.8)--(0.9,-1.4);
    \draw [->] (-0.8,-0.3)--(-1.4,-0.9);
    \node at (-0.3,1.3) {$p_1$};
    \node at (1.3,0.3) {$p_2$};   
    \node at (0.3,-1.3) {$p_3$};
    \node at (-1.3,-0.3) {$p_4$};
    \node at (-1.5,1.8) {$\mu_1,\; a_1$};
    \node at (-1.5,-1.8) {$\mu_4,\; a_4$};
    \node at (1.5,1.8) {$\mu_2,\; a_2$};
    \node at (1.5,-1.8) {$\mu_3,\; a_3$};
    \node at (2.5,0) {$=$};
    \node at (6.75,0.5) {
    \begin{minipage}{8cm}
    \begin{align*}
        g^2 & \left[ f^{a_1 a_2 b}f^{a_3 a_4 b}\left(g^{\mu_1 \mu_3}g^{\mu_2 \mu_4}-g^{\mu_1\mu_4}g^{\mu_2\mu_3}\right)\right. \\
         &+f^{a_1 a_3 b}f^{a_2 a_4 b}\left(g^{\mu_1 \mu_2}g^{\mu_3 \mu_4}-g^{\mu_1\mu_4}g^{\mu_2\mu_3}\right)\\
         &\left.+f^{a_1 a_4 b}f^{a_2 a_3 b}\left(g^{\mu_1 \mu_2}g^{\mu_3 \mu_4}-g^{\mu_1\mu_3}g^{\mu_2\mu_4}\right)\right].
    \end{align*}
    \end{minipage}
    };
\end{tikzpicture}
\end{center}
Meanwhile the propagator is modified to take the Proca form:
\begin{center}
\begin{tikzpicture}[scale=1, line width=1 pt]
    \draw[gluon] (0,0)--(2,0); 
    \draw[->] (0.5,0.5)--(1.5,0.5);
    \node at (1,0.8) {$p$};
    \node at (-0.5,0) {$\mu,\; a$};
    \node at (2.5,0) {$\nu,\; b$};
    \node at (3.7,0) {$=$};
    \node at (7.5,0.3) {
    \begin{minipage}{10cm}
    \begin{equation*}
        \frac{\delta^{ab}}{p^2+m^2}\left(g_{\mu\nu}+\frac{p_\mu p_\nu}{m^2}\right).
    \end{equation*}
    \end{minipage}
    };
    \node at (12,0) {};
\end{tikzpicture}
\end{center}

\section{4-point Graviton-Dilaton Amplitudes from Double-Copy}
\label{app:4amps}
The amplitudes given by the double-copy of massive Yang-Mills are given here:
	\be
	\begin{split}
		\label{eq:4gravamp}
		\mathcal{M}^{hhhh}_4&=-\frac{1}{4 M_p^2}\bigg(\frac{1}{m^2-2 p_{1 2}}\Big(-z_{1 4} z_{2 3} m^2+z_{1 3} z_{2 4} m^2+2 z_{1 2} z_{3 4}
		m^2+2 p_{1 2} z_{1 4} z_{2 3}\\
		&-2 p_{1 2} z_{1 3} z_{2 4}-2 p_{1 2} z_{1 2} z_{3 4}-4
		p_{1 3} z_{1 2} z_{3 4} +4 z_{3 4} zp_{1 3} zp_{2 1}-4 z_{3 4}
		zp_{1 2} zp_{2 3}\\
		&+4 z_{2 4} zp_{1 2} zp_{3 1}-4 z_{1 4}
		zp_{2 1} zp_{3 1}+4 z_{2 4} zp_{1 2} zp_{3 2} -4 z_{1 4}
		zp_{2 1} zp_{3 2}-4 z_{2 3} zp_{1 2} zp_{4 1}\\
		&+4 z_{1 3}
		zp_{2 1} zp_{4 1}+4 z_{1 2} zp_{3 2} zp_{4 1}-4 z_{2 3}
		zp_{1 2} zp_{4 2}+4 z_{1 3} zp_{2 1} zp_{4 2}-4 z_{1 2}
		zp_{3 1} zp_{4 2}\Big){}^2\\
        &+\left(2\leftrightarrow3\right)+\left(2\leftrightarrow4\right)\bigg).
	\end{split}
	\ee
	\be
	\begin{split}
		\mathcal{M}^{\phi\phi\phi\phi}_4&=\frac{1}{M_p^2}\bigg(-\frac{p_{1 3}^2 \left(75 m^2 p_{1 2}+34 p_{1 2}^2+116 m^4\right)}{72 m^4 \left(m^2-2
			p_{1 2}\right)}\\&+\frac{3}{64} \left(-24 m^2 p_{1 2}+48 p_{1 2}^2+115 m^4\right)
		\left(\frac{1}{2 p_{1 4}+m^2}+\frac{1}{2 p_{1 3}+m^2}\right)\\&+\frac{p_{1 3} \left(-41
			m^4 p_{1 2}-41 m^2 p_{1 2}^2-34 p_{1 2}^3+116 m^6\right)}{72 m^4 \left(m^2-2
			p_{1 2}\right)}\\&
		-\frac{-4751 m^4 p_{1 2}+744 m^2 p_{1 2}^2+368 p_{1 2}^3+3696 m^6}{288
			m^2 \left(m^2-2 p_{1 2}\right)}\bigg).
	\end{split}
	\ee
	\be
	\begin{split}
		\mathcal{M}_4^{h\phi\phi\phi}&=\frac{1}{6 \sqrt{3} m^4 M_p^2 \big(m^2-2 p_{1 2}\big) \big(m^2-2
			p_{1 3}\big) \big(m^2-2 \big(p_{1 4}\big)\big)} \big(m^{10} \big(19 zp_{1 2}^2-43 zp_{1 3} zp_{1 2}\\&+19 zp_{1 3}^2\big)-m^8 \big(p_{1 3}
		\big(136 zp_{1 2}^2-53 zp_{1 3} zp_{1 2}
		+76 zp_{1 3}^2\big)+p_{1 2} \big(76 zp_{1 2}^2-53
		zp_{1 3} zp_{1 2}\\&+136 zp_{1 3}^2\big)\big)+m^6 \big(p_{1 2}^2 \big(76 zp_{1 2}^2
		+61 zp_{1 3}
		zp_{1 2}+195 zp_{1 3}^2\big)+3 p_{1 3} p_{1 2} \big(39 zp_{1 2}^2
		\\&-53 zp_{1 3} zp_{1 2}+39
		zp_{1 3}^2\big)+p_{1 3}^2 \big(195 zp_{1 2}^2+61 zp_{1 3} zp_{1 2}+76
		zp_{1 3}^2\big)\big)\\&
		+m^4 \big(p_{1 2}^3 zp_{1 3} \big(10 zp_{1 2}+zp_{1 3}\big)+p_{1 3} p_{1 2}^2
		\big(-41 zp_{1 2}^2+10 zp_{1 3} zp_{1 2}
		-37 zp_{1 3}^2\big)\\&+p_{1 3}^2 p_{1 2} \big(-37
		zp_{1 2}^2+10 zp_{1 3} zp_{1 2}-41 zp_{1 3}^2\big)+p_{1 3}^3 zp_{1 2} \big(zp_{1 2}+10
		zp_{1 3}\big)\big)\\&
		+2 m^2 \big(p_{1 2}^4 zp_{1 3}^2
		-12 p_{1 3} p_{1 2}^3 zp_{1 2} zp_{1 3}-2
		p_{1 3}^2 p_{1 2}^2 \big(zp_{1 2}^2+13 zp_{1 3} zp_{1 2}+zp_{1 3}^2\big)\\&-12 p_{1 3}^3 p_{1 2}
		zp_{1 2} zp_{1 3}+p_{1 3}^4 zp_{1 2}^2\big)-4 p_{1 2} p_{1 3} \big(p_{1 2}+p_{1 3}\big) \big(p_{1 3}
		zp_{1 2}-p_{1 2} zp_{1 3}\big){}^2\big)\big).
	\end{split}
	\ee
	\be
	\begin{split}
		\mathcal{M}^{hh\phi\phi}_4&=\frac{1}{6 m^2 M_p^2
			\big(m^2-2 p_{1 2}\big) \big(m^2-2 p_{1 3}\big) \big(m^2-2
			p_{1 4}\big)} \Big(19 z_{1 2}^2 m^{10}\\&
		-z_{1 2} \big(62 p_{1 2} z_{1 2}+92 p_{1 3}
		z_{1 2}+zp_{1 2} zp_{2 1}-35 zp_{1 3} zp_{2 1}+17
		zp_{1 2} zp_{2 3}-18 zp_{1 3} zp_{2 3}\big)
		m^8\\& +\big(42 p_{1 2}^2 z_{1 2}^2+156 p_{1 3}^2 z_{1 2}^2+4 p_{1 3}
		\big(zp_{1 2} \big(zp_{2 1}+21 zp_{2 3}\big)-3
		zp_{1 3} \big(13 zp_{2 1}+6 zp_{2 3}\big)\big)
		z_{1 2}
		\\&+p_{1 2} \big(180 p_{1 3} z_{1 2}+zp_{1 2} \big(4 zp_{2 1}+31
		zp_{2 3}\big)-zp_{1 3} \big(121 zp_{2 1}+90
		zp_{2 3}\big)\big) z_{1 2}+34 zp_{1 3}^2
		zp_{2 1}^2\\&+zp_{1 2}^2 zp_{2 3}^2-33 zp_{1 2}
		zp_{1 3} zp_{2 3}^2+33 zp_{1 3}^2 zp_{2 1}
		zp_{2 3}-35 zp_{1 2} zp_{1 3} zp_{2 1}
		zp_{2 3}\big) m^6\\&+\big(13 z_{1 2}^2 p_{1 2}^3-4 z_{1 2} \big(14 p_{1 3}
		z_{1 2}+zp_{1 2} \big(zp_{2 1}-2
		zp_{2 3}\big)-zp_{1 3} \big(25 zp_{2 1}+27
		zp_{2 3}\big)\big) p_{1 2}^2\\&
		-\big(200 p_{1 3}^2 z_{1 2}^2+12 p_{1 3}
		\big(zp_{1 2} \big(zp_{2 1}+10 zp_{2 3}\big)-2
		zp_{1 3} \big(14 zp_{2 1}+9 zp_{2 3}\big)\big)
		z_{1 2}+zp_{1 2}^2 zp_{2 3}^2\\
		&-2 zp_{1 2} zp_{1 3}
		zp_{2 3} \big(34 zp_{2 1}+33 zp_{2 3}\big)+zp_{1 3}^2
		\big(100 zp_{2 1}^2+132 zp_{2 3} zp_{2 1}+33
		zp_{2 3}^2\big)\big) p_{1 2}\\
		&-p_{1 3} \big(128 p_{1 3}^2 z_{1 2}^2+4
		p_{1 3} \big(zp_{1 2} \big(zp_{2 1}+42 zp_{2 3}\big)-6
		zp_{1 3} \big(10 zp_{2 1}+3 zp_{2 3}\big)\big) z_{1 2}\\&
		+37
		zp_{1 2}^2 zp_{2 3}^2-2 zp_{1 2} zp_{1 3} zp_{2 3}
		\big(70 zp_{2 1}+33 zp_{2 3}\big)+zp_{1 3}^2 zp_{2 1}
		\big(103 zp_{2 1}+66 zp_{2 3}\big)\big) m^4\\&-2
		\big(z_{1 2}^2 p_{1 2}^4+2 z_{1 2} \big(9 p_{1 3} z_{1 2}-zp_{1 3}
		zp_{2 1}+zp_{1 2} zp_{2 3}\big) p_{1 2}^3+\big(-\big(32
		zp_{2 1}^2+66 zp_{2 3} zp_{2 1}\\&
		+33 zp_{2 3}^2\big)
		zp_{1 3}^2-2 zp_{1 2} zp_{2 1} zp_{2 3}
		zp_{1 3}+zp_{1 2}^2 zp_{2 3}^2+2 p_{1 3} z_{1 2}
		\big(zp_{1 2} \big(7 zp_{2 3}-2
		zp_{2 1}\big)\\&
		+zp_{1 3} \big(29 zp_{2 1}+36
		zp_{2 3}\big)\big)\big) p_{1 2}^2-2 p_{1 3} \big(24 p_{1 3}^2
		z_{1 2}^2+p_{1 3} \big(zp_{1 2} \big(2 zp_{2 1}+27
		zp_{2 3}\big)\\&-9 zp_{1 3} \big(7 zp_{2 1}+4
		zp_{2 3}\big)\big) z_{1 2}+33 zp_{1 3}
		\big(zp_{2 1}+zp_{2 3}\big) \big(zp_{1 3}
		zp_{2 1}-zp_{1 2} zp_{2 3}\big)\big) p_{1 2}\\&
		-p_{1 3}^2
		\big(32 p_{1 3}^2 z_{1 2}^2-68 p_{1 3} \big(zp_{1 3}
		zp_{2 1}-zp_{1 2} zp_{2 3}\big) z_{1 2}+35
		\big(zp_{1 3} zp_{2 1}-zp_{1 2}
		zp_{2 3}\big){}^2\big)\big) m^2\\&+4 p_{1 2} p_{1 3}
		\big(p_{1 2}+p_{1 3}\big) \big(p_{1 2} z_{1 2}+2 p_{1 3} z_{1 2}-zp_{1 3}
		zp_{2 1}+zp_{1 2} zp_{2 3}\big){}^2\big)\Big).
	\end{split}
	\ee
	\be
	\begin{split}
		\mathcal{M}_4^{hhh\phi}&=
		\frac{-1}{2 \sqrt{3}
			M_p^2 \big(m^2-2 p_{1 2}\big) \big(m^2-2 p_{1 3}\big) \big(m^2-2
			(p_{1 4}\big)} \big(z_{1 2} z_{1 3} z_{2 3} m^8-\big(-14 zp_{3 1}^2 z_{1 2}^2
		\\&-2
		zp_{3 2}^2 z_{1 2}^2-8 zp_{3 1} zp_{3 2} z_{1 2}^2+4 p_{1 2}
		z_{1 3} z_{2 3} z_{1 2}+4 p_{1 3} z_{1 3} z_{2 3} z_{1 2}-7 z_{2 3} zp_{1 2}
		zp_{3 1} z_{1 2}\\&+13 z_{2 3} zp_{1 3} zp_{3 1} z_{1 2}+13 z_{1 3}
		zp_{2 1} zp_{3 1} z_{1 2}-7 z_{1 3} zp_{2 3} zp_{3 1}
		z_{1 2}+z_{2 3} zp_{1 2} zp_{3 2} z_{1 2}\\&+5 z_{2 3} zp_{1 3}
		zp_{3 2} z_{1 2}-7 z_{1 3} zp_{2 1} zp_{3 2} z_{1 2}-11 z_{1 3}
		zp_{2 3} zp_{3 2} z_{1 2}+z_{2 3}^2 zp_{1 2}^2+z_{2 3}^2
		zp_{1 3}^2\\&
		-14 z_{1 3}^2 zp_{2 1}^2-2 z_{1 3}^2 zp_{2 3}^2+10
		z_{2 3}^2 zp_{1 2} zp_{1 3}
		+13 z_{1 3} z_{2 3} zp_{1 2}
		zp_{2 1}-7 z_{1 3} z_{2 3} zp_{1 3} zp_{2 1}\\&
		+5 z_{1 3} z_{2 3}
		zp_{1 2} zp_{2 3}+z_{1 3} z_{2 3} zp_{1 3} zp_{2 3}-8
		z_{1 3}^2 zp_{2 1} zp_{2 3}\big) m^6+\big(4 z_{1 2} z_{1 3} z_{2 3}
		p_{1 2}^2\\&
		+\big(-41 zp_{3 1}^2 z_{1 2}^2-5 zp_{3 2}^2 z_{1 2}^2-14
		zp_{3 1} zp_{3 2} z_{1 2}^2+12 p_{1 3} z_{1 3} z_{2 3} z_{1 2}+40
		z_{1 3} zp_{2 1} zp_{3 1} z_{1 2}
		\\&-28 z_{1 3} zp_{2 3}
		zp_{3 1} z_{1 2}-16 z_{1 3} zp_{2 1} zp_{3 2} z_{1 2}-20 z_{1 3}
		zp_{2 3} zp_{3 2} z_{1 2}-44 z_{1 3}^2 zp_{2 1}^2-20 z_{1 3}^2
		zp_{2 3}^2
		\\&+z_{2 3}^2 \big(zp_{1 2}^2+22 zp_{1 3}
		zp_{1 2}-11 zp_{1 3}^2\big)-32 z_{1 3}^2 zp_{2 1}
		zp_{2 3}+4 z_{2 3} \big(z_{1 3} \big(zp_{1 2} \big(7
		zp_{2 1}+2 zp_{2 3}\big)
		\\&+zp_{1 3} \big(4 zp_{2 3}-7
		zp_{2 1}\big)\big)+z_{1 2} \big(zp_{1 2} \big(zp_{3 2}-4
		zp_{3 1}\big)+zp_{1 3} \big(13 zp_{3 1}+2
		zp_{3 2}\big)\big)\big)\big) p_{1 2}
		\\&+p_{1 3} \big(-44
		zp_{3 1}^2 z_{1 2}^2-20 zp_{3 2}^2 z_{1 2}^2-32 zp_{3 1}
		zp_{3 2} z_{1 2}^2+4 p_{1 3} z_{1 3} z_{2 3} z_{1 2}+40 z_{1 3}
		zp_{2 1} zp_{3 1} z_{1 2}
		\\&-16 z_{1 3} zp_{2 3} zp_{3 1}
		z_{1 2}-28 z_{1 3} zp_{2 1} zp_{3 2} z_{1 2}-20 z_{1 3} zp_{2 3}
		zp_{3 2} z_{1 2}-41 z_{1 3}^2 zp_{2 1}^2-5 z_{1 3}^2
		zp_{2 3}^2
		\\&+z_{2 3}^2 \big(-11 zp_{1 2}^2+22 zp_{1 3}
		zp_{1 2}+zp_{1 3}^2\big)-14 z_{1 3}^2 zp_{2 1}
		zp_{2 3}+4 z_{2 3} \big(z_{1 3} \big(zp_{1 3} \big(zp_{2 3}-4
		zp_{2 1}\big)\\&
		+zp_{1 2} \big(13 zp_{2 1}+2
		zp_{2 3}\big)\big)+z_{1 2} \big(zp_{1 3} \big(7 zp_{3 1}+2
		zp_{3 2}\big)+zp_{1 2} \big(4 zp_{3 2}-7
		zp_{3 1}\big)\big)\big)\big)\big) m^4\\&
		+2 \big(\big(13
		zp_{3 1}^2 z_{1 2}^2+zp_{3 2}^2 z_{1 2}^2-2 zp_{3 1}
		zp_{3 2} z_{1 2}^2-4 p_{1 3} z_{1 3} z_{2 3} z_{1 2}-14 z_{1 3}
		zp_{2 1} zp_{3 1} z_{1 2}
		\\&+14 z_{1 3} zp_{2 3} zp_{3 1}
		z_{1 2}+2 z_{1 3} zp_{2 1} zp_{3 2} z_{1 2}-2 z_{1 3} zp_{2 3}
		zp_{3 2} z_{1 2}+16 z_{1 3}^2 zp_{2 1}^2+16 z_{1 3}^2
		zp_{2 3}^2\\&
		-2 z_{2 3} \big(z_{1 3} \big(zp_{1 2}-7
		zp_{1 3}\big) \big(zp_{2 1}-zp_{2 3}\big)+
		z_{1 2}zp_{1 3} \big(13 zp_{3 1}-zp_{3 2}\big)\\&+z_{1 2}zp_{1 2}
		\big(zp_{3 2}-zp_{3 1}\big)\big)+z_{2 3}^2 \big(zp_{1 2}^2-2 zp_{1 3}
		zp_{1 2}+13 zp_{1 3}^2\big)+16 z_{1 3}^2 zp_{2 1}
		zp_{2 3}\big) p_{1 2}^2
		\\&-2 p_{1 3}
		\big(2 p_{1 3} z_{1 2} z_{1 3} z_{2 3}-3 \big(\big(5 zp_{3 1}^2+2
		zp_{3 2} zp_{3 1}+zp_{3 2}^2\big) z_{1 2}^2+z_{1 3} \big(3
		zp_{2 3} \big(zp_{3 1}+zp_{3 2}\big)
		\\&+zp_{2 1} \big(3
		zp_{3 2}-5 zp_{3 1}\big)\big) z_{1 2}-4 z_{2 3}^2 zp_{1 2}
		zp_{1 3}+z_{1 3}^2 \big(5 zp_{2 1}^2+2 zp_{2 3}
		zp_{2 1}+zp_{2 3}^2\big)
		\\&-z_{2 3} \big(z_{1 3} \big(zp_{1 3}
		\big(zp_{2 3}-3 zp_{2 1}\big)+zp_{1 2} \big(5
		zp_{2 1}+zp_{2 3}\big)\big)
		+z_{1 2} \big(zp_{1 2}
		\big(zp_{3 2}-3 zp_{3 1}\big)\\&
		+zp_{1 3} \big(5
		zp_{3 1}+zp_{3 2}\big)\big)\big)\big)\big) p_{1 2}+p_{1 3}^2
		\big(16 \big(zp_{3 1}^2+zp_{3 2}
		zp_{3 1}+zp_{3 2}^2\big) z_{1 2}^2\\&
		-2 z_{1 3} \big(7
		zp_{2 1}-zp_{2 3}\big) \big(zp_{3 1}-zp_{3 2}\big)
		z_{1 2}+z_{2 3}^2 \big(13 zp_{1 2}^2-2 zp_{1 3}
		zp_{1 2}
		+zp_{1 3}^2\big)\\&+z_{1 3}^2 \big(13 zp_{2 1}^2-2
		zp_{2 3} zp_{2 1}+zp_{2 3}^2\big)-2 z_{2 3} \big(z_{1 3}
		\big(zp_{1 2} \big(13 zp_{2 1}-zp_{2 3}\big)
		\\&+zp_{1 3}
		\big(zp_{2 3}-zp_{2 1}\big)\big)-z_{1 2} \big(7
		zp_{1 2}-zp_{1 3}\big)
		\big(zp_{3 1}-zp_{3 2}\big)\big)\big)\big) m^2
		\\&-4 p_{1 2}
		p_{1 3} \big(p_{1 2}+p_{1 3}\big) \big(z_{2 3}
		\big(zp_{1 2}-zp_{1 3}\big)+z_{1 3}
		\big(zp_{2 3}-zp_{2 1}\big)+z_{1 2}
		\big(zp_{3 1}-zp_{3 2}\big)\big){}^2\big).
	\end{split}
	\ee

	\section{BCJ Relations as Null Vectors}
	\label{app:BCJnull}
	The BCJ relations can be obtained as null space relations of the matrix of bi-adjoint scalar amplitudes. To show this, one must first notice a remarkable property about these amplitudes. Just as in the massless case, bi-adjoint scalar theory acts as an identity for the massive double-copy\footnote{It is an interesting fact that this is true whether or not the spectral conditions hold.}, i.e.
	\begin{align}
		A\otimes\text{BS} =A\,.
	\end{align}
	
	To express this in matrix notation, let us first choose an $(n-2)!$ DDM basis. From this, we choose BCJ-independent $(n-3)!$ sub-bases $\alpha$, $\beta$ and $\gamma$ and use the KLT formula,
	\begin{align}
		\mathcal{A}^{\phi^3}[\alpha|\beta]\ \mathcal{A}^{\phi^3}[\beta|\gamma]^{-1}\ \vec{\mathcal{A}}^A[\gamma]&=\vec{\mathcal{A}}^A[\alpha].
	\end{align}
	The BCJ relations are consistency conditions that make the KLT formula basis-independent. For example, consider another $(n-3)!$ sub-basis $\tilde{\gamma}$. We can then express a BCJ relation as
	\begin{align}
		\mathcal{A}^{\phi^3}[\alpha|\beta]\ \mathcal{A}^{\phi^3}[\beta|\gamma]^{-1}\ \vec{\mathcal{A}}^A[\gamma]&=\mathcal{A}^{\phi^3}[\alpha|\beta]\ \mathcal{A}^{\phi^3}[\beta|\tilde{\gamma}]^{-1}\ \vec{\mathcal{A}}^A[\tilde{\gamma}]\,.
	\end{align}
	
	We now embed these matrices in our original $(n-2)!$ DDM basis. The matrix $\mathcal{A}^{\phi^3}[\alpha|\beta]$ is padded with the remaining bi-adjoint scalar amplitudes, while we pad the vector
	\be
	\left(\mathcal{A}^{\phi^3}[\beta|\gamma]^{-1}\vec{\mathcal{A}}^A[\gamma]-\mathcal{A}^{\phi^3}[\beta|\tilde{\gamma}]^{-1}\vec{\mathcal{A}}^A[\tilde{\gamma}]\right)
	\ee
	with zeros. This gives us the following null vector equation,
	\begin{align}
	\mathcal{A}^{\phi^3}[\alpha|\beta]\left(\mathcal{A}^{\phi^3}[\beta|\gamma]^{-1}\vec{\mathcal{A}}^A[\gamma]-\mathcal{A}^{\phi^3}[\beta|\tilde{\gamma}]^{-1}\vec{\mathcal{A}}^A[\tilde{\gamma}]\right) &=0\,.
	\end{align}
	
	To connect this to the BCJ relations of theory $A$, we consider a double-copy of $A$ with itself,
	\begin{align}
		A\otimes A &= B.
	\end{align}
	Choosing the same sub-bases as previously, we can rewrite the KLT formula,
	\begin{align}
	\vec{\mathcal{A}}^A[\beta]^T\ \mathcal{A}^{\phi^3}[\beta|\gamma]^{-1}\ \vec{\mathcal{A}}^A[\gamma]&=\vec{\mathcal{A}}^B.
	\end{align}
	Again the BCJ relations are given by demanding basis-independence of this formula,
	\begin{align}
	&\vec{\mathcal{A}}^A[\beta]^T\ \mathcal{A}^{\phi^3}[\beta|\gamma]^{-1}\ \vec{\mathcal{A}}^A[\gamma]=\vec{\mathcal{A}}^A[\beta]^T\ \mathcal{A}^{\phi^3}[\beta|\tilde{\gamma}]^{-1}\ \vec{\mathcal{A}}^A[\tilde{\gamma}]\nonumber\\
	\Rightarrow\ \ &\vec{\mathcal{A}}^A[\beta]^T\left(\mathcal{A}^{\phi^3}[\beta|\gamma]^{-1}\vec{\mathcal{A}}^A[\gamma]-\mathcal{A}^{\phi^3}[\beta|\tilde{\gamma}]^{-1}\vec{\mathcal{A}}^A[\tilde{\gamma}]\right)=0.
	\end{align}
	
	We recognize this vector as being a null vector of $\mathcal{A}^{\phi^3}[\alpha|\beta]$. Indeed this equation must hold for all choices of $\gamma$ and $\tilde{\gamma}$. At 4- and 5-point, we observe that different choices of $\gamma$ and $\tilde{\gamma}$ span the null space of $\mathcal{A}^{\phi^3}[\alpha|\beta]$, allowing us to generalize this equation to,
	\begin{align}
	\label{eq:nullvec}
	\vec{\mathcal{A}}^A[\beta]\cdot \vec{n}=0,
	\end{align}
	where the vector $\vec{n}$ is any null vector of the matrix of bi-adjoint scalar amplitudes.
	
	Thus \eqref{eq:nullvec} is an equivalent representation of the BCJ relations. Since the number of null vectors of $\mathcal{A}^{\phi^3}[\alpha|\beta]$ is exactly the number of independent BCJ relations, we expect this equivalence to continue to any $n$-point. 

\bibliographystyle{JHEP}
\bibliography{MassiveDoubleCopy.bib}

\end{document}